\documentclass[10pt]{article}
\usepackage{amssymb,dsfont,amsmath,amsfonts,bm,mathabx}
\usepackage[active]{srcltx}
\usepackage{slashed}
\usepackage{color}
\usepackage{graphicx}
\usepackage{cite}

\advance \textheight by \topskip
 \def\*{{\varstar}}
 \def\cL{\mathcal{L}}

\def\be{\begin{equation}}
\def\ee{\end{equation}}
\def\bea{\begin{eqnarray}}
\def\eea{\end{eqnarray}}

\def\a{\alpha}
\def\b{\beta}

\def\e{\epsilon}

\def\bfcL{{\boldsymbol{\mathcal{L}}}}
\def\bfa{{\boldsymbol{\alpha}}}
\def\bfb{{\boldsymbol{\beta}}}
\def\bfxi{{\boldsymbol{\xi}}}
\def\bfz{{\boldsymbol{\zeta}}}
\def\bfC{{\boldsymbol{C}}}
\def\bmu{{\boldsymbol{\mu}}}
\def\bnu{{\boldsymbol{\nu}}}
\def\blambda{{\boldsymbol{\lambda}}}
\def\brho{{\boldsymbol{\rho}}}
\def\bxi{{\boldsymbol{\rho}}}

\def\bL{{\boldsymbol{\mathcal{L}}}}
\def\ba{{\boldsymbol{\alpha}}}
\def\bb{{\boldsymbol{\beta}}}

\def\fr{{\mathfrak{r}}}
\def\ft{{\mathfrak{t}}}
\def\ap{{\overrightarrow{p}}}
\def\aq{{\overrightarrow{q}}}
\def\cd{{\cdot}}
\def\av{{\overrightarrow{v}}}

\def\bZ{{\boldsymbol{Z}}}

\def\bLam{{\bf \Lambda}}
\def\bV{{\bf V}}
\def \bY{{\bf Y}}

\def\ssl{{\slashed{\ell}_\texttt{H}}}

\def\bX{\mathbf{X}}

\def\ssl{\slashed{\ell}_\texttt{HP}}

\newcommand{\bss}{\boldsymbol}

\begin{document}

\title{From phase space to multivector matrix models}

\author{\sf  Mauricio Valenzuela\footnote{\texttt{valenzuela.u} \textit{at} \texttt{gmail.com}}
\\[5mm]
{\textit{Facultad de Ingenier\'ia y Tecnolog\'ia}}\\
{\textit{Universidad San Sebasti\'an, General Lagos 1163, Valdivia 5110693, Chile}}
}
\date{}

\maketitle

\begin{abstract}
 
{ Combining elements of twistor-space, phase space and Clifford algebras, we propose a framework for the construction and quantization of certain  (quadric)  varieties described by Lorentz-covariant multivector coordiantes.}
The correspondent multivectors  can be parametrized by second order polynomials in the phase space. Thus the multivectors play a double role, as covariant objects in $D=2,3,4 \texttt{ Mod } 8$ space-time dimensions, and as mechanical observables of a non-relativistic system in $2^{[D/2]-1}$ euclidean dimensions. The latter attribute permits a dual interpretation of concepts of non-relativistic  mechanics as applying to relativistic space-time geometry. Introducing the Groenewold-Moyal $\*$-product and Wigner distributions in phase space induces Lorentz-covariant non-commutativity and it provides the spectra of geometrical observables. { We propose also new (multivector) matrix models, interpreted as descending from the interaction term of a Yang-Mills theory with minimally coupled massive fermions, in the large-$N$ limit, which serves as a physical model containing the constructed multivector (fuzzy) geometries.  We also include a section on speculative aspects on a possible cosmological effect and the origin of space-time entropy.}
 
\end{abstract}

\newpage

\tableofcontents

\section{Introduction}

It is expected that at scales of the order of the Planck length the theoretical description of the structure of the space-time  must combine aspects of geometry and quantum mechanics. Black holes information (loss) paradox  (see e.g. \cite{Giddings:1995gd,Mathur:2009hf}) and AdS/CFT correspondence \cite{Maldacena:1997re,Witten:1998qj} have motivated new ways to look at the problem of quantum gravity \cite{Hooft:1993gx,Thorn:1991fv,Susskind:1994vu,Bousso:2002ju} (see also \cite{Sakharov:1967pk,Visser:2002ew}). In \cite{Verlinde:2010hp} is pointed out that gravity may appear as an entropic (reaction) force, a product of distributions of information/entropy in space. Since entropy often appears from phase space probability distributions, the latter point of view suggests that there should be a hidden phase space from which entropy does emerge. In this spirit, we present here a construction in which phase spaces of several dimensions appear as underlying { multivector spaces, i.e. whose coordinates are labeled by antisymmetric Lorentz indices}. This permits the application of concepts and tools of non-relativistic mechanics  to space-time geometries in a novel way. Indeed, we shall extend the notion of twistor-space \cite{Penrose:1967wn,Penrose:1972ia} to phase spaces of several dimensions such that we will be able to construct relativistic geometries beyond the light-cone. To this notion of phase space we shall refer as to twistor--phase space ({\bf TPS}).

The inclusion of $D$-dimensional Lorentz algebras in the algebra of symplectic transformations $so(D-1,1) \hookrightarrow sp(2^{[D/2]})$ induces the action of Lorentz transformations on functions in phase space. Among these functions there will be sets of them transforming as anti-symmetric multivectors of the form $x^{\mu[n]}:= x^{[\mu_1 \cdots \mu_n]}$, $\mu=0,1, \dots, D-1$, for several $n$'s.
 Complementing these classes there will appear also Wigner distributions in phase space. We shall identify the rank-$1$ tensors as  coordinates of events in Minkowski space { which together with higher order tensors will span coordinates of an algebraic multivector variety of dimension $2^{[D/2]}$, to be denoted $\bX[\mathcal{M}_D]$, $D=2,3,4 \texttt{ mod } 8$. In general, the single-index components ($x^\mu$'s) of these coordinates will label points on subspaces of the Minkowski space, on the light-cone for  $D=2+1,\, 3+1$, and in the non-negative--time sub-space $\mathds{R}^{\geq 0} \times \mathds{R}^{D-1}$ in higher dimensions. We shall elaborate more on this in the subsequent sections.}  

We quantize the space $\bX[\mathcal{M}_D]$ using the \textit{phase space  formulation of quantum mechanics}, i.e. in the framework of \textit{deformation quantization} \cite{Groenewold:1946kp,Moyal:1949sk,Fairlie:1964qt,Bayen:1977ha,Bayen:1977hb,Fairlie:1991yt,Kontsevich:1997vb,Zachos:2001ux}. As a result we obtain a non-commutative geometry which generalizes the Snyder space \cite{Snyder:1946qz} and other non-commutative extensions of space-time (cf. \cite{Connes:1994yd}).  The non-commutative multivector space constructed so far, denoted $\bX[\mathcal{M}_D\,,\*]$, is extended with Wigner distributions which will allow us to measure the mean values of functions in TPS, and of functionals of multivectors. In this formalism the product of functions in phase space is given by the $\*$-product, a non-local convolution-like product of functions which respect symplectic and Lorentz covariance. The quantization constant, referred as to \textit{Heisenberg-Planck length}, enters as the central charge of the Heisenberg algebra of TPS.

A natural question is, What is underlying model for the non-commutative geometry $\bX[\mathcal{M}_D, \*]$? We shall propose a matrix model in zero dimensions analogous to the IKKT model \cite{Ishibashi:1996xs}  with a multivector target space. We shall see that $\bX[\mathcal{M}_D, \*]$ appears as solution of this matrix model. We show alsto that these geometries enjoy (infinite) higher-spin-gravity--type of symmetries  \cite{Vasiliev:2004cm}. 

Applying concepts of thermodynamics, we shall observe that the volume/entropy (for constant statistical distribution) of a solid ball in $2^{[D/2]}$-dimensional TPS is proportional to the area of a $2^{[D/2]-1}$-sphere. The relation is holographic for $D=3,4$, i.e. the respective spheres  are $S^1$ and $S^2$ but it is projective for $D>4$, i.e. where the entropy is proportional to a sphere of dimension higher than the space-time.

We shall see that in  $D=3+1$,  $\bX[\mathcal{M}_{4}]$ is equivalent to a subvariety of the Grassmannian $\texttt{Gr}(2,4)$ since the correspondent rank-2 tensor coordinates are  Pl\"ucker coordinates, i.e. they label planes through the origin of the Minkowski space-time, are generated by at least one light-ray. We shall also compute the entropy of a solid ball in four dimensional TPS and observe that for a constant probability distribution  it is  proportional to the area of a 2-sphere immersed in $D=3+1$ space-time. This observation suggests that the origin of the holographic principle is consequence of twistor space as the underlying structure of space-time.

In summary, we present a basis for the development of a theory in which the space-time appears as a mechanical observable in TPS,  where the tools of mechanics (classical, quantum and statistical) can be used in a novel way to derive statistical/information properties of the space-time. We expect that our approach will be useful to elaborate new frameworks for quantum gravity along similar lines than Verlinde's ideas \cite{Verlinde:2010hp}.

 \subsection{Plan of the paper} 

\begin{itemize}

\item In section \ref{PSQ} we review some elements of phase space formulation of quantum mechanics \cite{Groenewold:1946kp,Moyal:1949sk,Fairlie:1964qt,Zachos:2001ux,Fairlie:1991yt}. The Groenewold-Moyal $\*$-product is defined and we solve the Harmonic oscillator problem as an useful example.

\item In section \ref{TPS} we introduce the twistor--phase space. Using Clifford algebras we construct the multivector coordinates of the variety $\bX[\mathcal{M}_D]$ in terms of second order homogeneous polynomials in phase space. We identify the algebraic constraints satisfied by these coordinates. 

\item  In section \ref{QinTPS} we quantize $\bX[\mathcal{M}_D]$ in TPS \textit{\`a la} Groenewold-Moyal and we introduce the {geometrical Wigner distribution}. We compute the spectrum of some observables such as, the time coordinate and the area/volume of spheres. 
 In the subsection \ref{HSS} we show that the non-commutative version of $\bX[\mathcal{M}_D]$, i.e. $\bX[\mathcal{M}_D,\*]$, enjoy {higher spin symmetries}.
 
\item  In section \ref{MVMM} we construct a new multivector matrix model, from which $\bX[\mathcal{M}_D,\*]$ appears as solution.
 
  \item In section \ref{NCS}  we present some ideas on physical implications of non-commutative light-cones and a possible origin of space-time entropy.

\item Section \ref{CONC} we present our conclusions.

\end{itemize}

\section{Elements of the phase space formulation of quantum mechanics}\label{PSQ}

 Quantization in Hilbert spaces assigns operators to functions on the phase space provided a prescription of ordering. In contrast, in deformation quantization a $\*$-product is introduced which deforms the pointwise product of functions, thus implying  non-commutativity of phase space. The {Wigner distribution} is introduced to measure the expectation values of observables, and it is obtained as solution of the $\*$-genvalue problem \cite{Fairlie:1964qt,Fairlie:1991yt}.  In what follows we will present some elements of this approach which are necessary for our main task. The reader may use references \cite{Bayen:1977ha,Bayen:1977hb,Curtright:2011vw,Fairlie:1991yt}  for further details. 
 
Consider a $2d$-dimensional phase space, spanned by coordinates, 
\be \label{Ln}
\bfcL_{\bfa}=(q_1,\cdots q_d,p_1,\cdots,p_d)\, \quad \in \quad \mathcal{P}^{2d}\,,
\ee
 composed of $d$ positions ($q$'s) and $d$ momenta ($p$'s). 
For two arbitrary functions $f$ and $g$,  the Groenewold-Moyal $\*$-product is defined as,
\be \label{f*g2n}
 f({\bfcL})\* g({\bfcL})~=~\int \frac{d^{2d}{\bfxi} \, d^{2d}{\bfz}}{(\pi{\ssl})^{2d}} \, \exp\Big( -\tfrac{2i}{{\ssl}}\, \bar{\bfxi} \bfz \Big) \, f({\bfcL}+\bfxi )\, g({\bfcL}+\bfz)\,, 
\ee
where $\bar{\bfxi} \bfz = \bfxi^{\bfa} \bfz_{\bfa}$, ${\ssl}$ is the quantum (Planck scale) deformation parameter, 
and we have used the symplectic matrix 
\begin{eqnarray}
\bfC_{\bfa \bfb}=\bfC^{\bfa \bfb}:=
\left(%
\begin{array}{cc}
  0 & \mathds{1}_{d\times d} \\
  -\mathds{1}_{d\times d} & 0 \\
\end{array}
\label{sM1} \right) \,,
\end{eqnarray}
to rise the phase space index according to the following conventions,
\be \label{sM2}
\bfxi^\bfa = \bfC^{\bfa \bfb} \bfxi_\bfb \, , \qquad \bfxi_\bfa = \bfxi^\bfb \bfC_{\bfb \bfa}\, , \quad \hbox{where}\quad \bfC_{\bfa \bfb} = \bfC^{\bfa \bfb}\,.
\ee 
{ Note that we use a bar for phase space coordinates with the index up, e.g. $\bar{\bfcL} {}^\bfa:=\bfcL^\bfa=\bfC^{\bfa \bfb}\bfcL_\bfb$. }
The complex conjugation acts on the $\*$-product of two functions $f$ and $g$ as an anti-authomorphism,
\be \label{fg*}
(f \* g )^*=g^* \* f^* \,.
\ee

The $\*$-product of two vectors yields, 
\be \label{L*L4}
\bfcL_\bfa \* \bfcL_\bfb = \bfcL_\bfa  \bfcL_\bfb + \frac{i {\ssl}}{2} \bfC_{\bfa \bfb}\,,
\ee 
where the symmetric part, $\bfcL_\bfa  \bfcL_\bfb=\bfcL_\bfb  \bfcL_\bfa$ is just the ``classical product" of commuting variables. For any function $f$ and a vector $\bfcL_\bfa$, 
\be \label{L*f}
\bfcL_\bfa \* f(\cL) = \bfcL_\bfa  f(\cL) + \frac{i {\ssl} }{2} \frac{\partial f(\bfcL)}{\partial \bfcL^\bfa}\,,\qquad
 f(\cL) \* \bfcL_\bfa = \bfcL_\bfa  f(\cL) - \frac{i {\ssl} }{2} \frac{\partial f(\bfcL)}{\partial \bfcL^\bfa}\, .
\ee
The $\*$-commutator is defined as
\be \label{*comm}
[ f \,,  g \, ]_\*:= f \, \*\, g - g \, \*\, f \,.
\ee
From \eqref{L*f} we obtain, 
\be \label{comdif}
[\bfcL_\bfa ,\,  f(\cL)]_{\*} =i {\ssl} \, \frac{\partial f(\bfcL)}{\partial \bfcL^\bfa}\,.
\ee
We define the $\*$-anti-commutator, 
\be \label{*anticomm}
\{ f \,,  g \, \}_\*:= f \, \*\, g + g \, \*\, f \,.
\ee
Note that if a phase-space vector is involved, the anti-commutator extracts the classical  component of the product \eqref{L*f},
\be \label{anticomdif}
\{\bfcL_\bfa ,\,  f(\cL)\}_{\*} = 2 \bfcL_\bfa  \, f(\cL) =2  f(\cL) \, \bfcL_\bfa  \,.
\ee

The Poisson bracket is obtained in the limit
\be \label{poisson}
\{f \,,  g \}_{\texttt{PB}}:= \lim_{\ssl \rightarrow 0} \frac{ [f \,,  g  ]_\*}{i \ssl}.
\ee

Fully symmetric product of phase space vectors,
\be 
\bfcL_{(\bfa_1}\*  \bfcL_{\bfa_2}\* \cdots \* \bfcL_{\bfa_n)}:=\frac{1}{n!}\,  \sum_{\sigma} \bfcL_{\sigma(\bfa_1)}\*  \bfcL_{\sigma(\bfa_2)} \* \cdots \* \bfcL_{\sigma(\bfa_n)}  \,, \nonumber 
\ee
where $\sigma$ belongs to the permutation group of $n$ elements,
 do not generate quantum corrections;
\be\label{a*n}
\bfcL_{(\bfa_1}\*  \bfcL_{\bfa_2}\* \cdots \* \bfcL_{\bfa_n)}= \bfcL_{\bfa_1}  \bfcL_{\bfa_2}\cdots \bfcL_{\bfa_n}  \,,
\ee
i.e. the r.h.s. is just a real-valued monomial in the phase space. 

It is convenient to use the following notation for the product \eqref{a*n},
\be \label{lan}
\bfcL_{\bfa(n)}:=\bfcL_{\bfa_1}  \bfcL_{\bfa_2}\cdots \bfcL_{\bfa_n}\,.
\ee 
The algebra of these generators, with the $\*$-product, span a representation of the associative Weyl algebra{, denoted ${\cal U}_{2d}$,} i.e. the universal enveloping algebra of the Heisenberg algebra  
\be \label{HA}
[ \bfcL_\bfa  \,, \,\bfcL_\bfb \, ]_\*:=  {i {\ssl}} \,\bfC_{\bfa \bfb} \,,
\ee
with central charge $\ssl$.
 The second order generators
\be \label{sp2d} 
\bfcL_{(\bfa_1} \*  \bfcL_{\bfa_2)}=\bfcL_{\bfa_1}  \bfcL_{\bfa_2}\,,
\ee
with the Moyal bracket \eqref{*comm} product, generate a representation of the $sp(2d)$ algebra.

An essential element of quantization in phase space is the Wigner  distribution, denoted by $W=W(q,p)=W(\bfcL)$, which represents the state of a physical system with respect to the Hamiltonian function. The Wigner distribution is not necessarily positive definite, it is not a \textit{bona fide} probability distribution but it has a probabilistic meaning. The reader may consult reference \cite{18757248} for a didactic explanation of the physical meaning of the Wigner function and of its experimental observation. 

The uncertainty of position and momentum follows from non-commutativity in phase-space \eqref{HA} (see proof in \cite{Fairlie:1991yt} or \cite{Zachos:2001ux}),
$$
\Delta q \Delta
p \geq \frac{{\ssl}}2 \,.$$

The Wigner distribution is normalized such that,
\be \label{Wignereq}
\int d^{2n}\bL \; \; W =   1 \,,
\ee 
and the expectation value of an observable $f(\bfcL)$ is given by the integral,
\be\label{expect}
\langle f(\bfcL) \rangle = \int d^{2n}\bL \, \, f(\bfcL) \, \* \, W(\bfcL)  \, . 
\ee
 For pure states the Wigner distribution can be expressed in terms of wave functions \cite{Wigner:1932eb}. For instance, in the one dimensional case,
\be \label{Wpsi}
W(q,p)= \frac{1}{\pi {\ssl}} \int  \, ds \, \psi_q(q+s) \psi_q^*(q-s) \, \exp ({2i s p}/{{\ssl}} ) = \frac{1}{\pi {\ssl}} \int \, ds \, \psi_p(p+s) \psi_p^*(p-s) \, \exp ({2 i s p}/{{\ssl}} ) \, . 
\ee
 The marginal probabilities in coordinate or momentum space are respectively given by the \textit{shadows},
\be \label{shadowspsi}
|\psi_q(q)|^2= \int \,  dp  \, W(q,p) \, ,\qquad |\psi_p(p)|^2= \int  \, dq \, W(q,p)   \, . 
\ee
Non-negative expectation values are obtained for expressions of the form (see the proof e.g. in \cite{Zachos:2001ux,Curtright:2011vw}),
\be
\langle f^* (\bfcL) \* f(\bfcL) \rangle \; \geq \; 0 \,.
\ee 
 A function $f(\lambda)$ can be promoted to a $\*$-function replacing its Taylor expansion
 $$
 f(\lambda)= \sum_{l}\, f_l \; \lambda^l\,,
 $$
 for constant $f_l$'s, by
 \begin{eqnarray}\label{*T}
   \begin{array}{cc}
f_\*(\lambda) := \sum_{l} \, f_l & {\underbrace{\lambda\* \cdots \* \lambda}} \, .\\  
 & l-\hbox{times} \end{array} 
 \end{eqnarray}
In particular, a Weyl ordered $\*$-function  
$$
[f_\* (\bfcL)]_W:=\sum_{\bfa(n)} \, f^{\bfa(n)} \, \bfcL_{(\bfa_1}\*  \bfcL_{\bfa_2}\* \cdots \* \bfcL_{\bfa_n)}\,,
$$
where $f^{\bfa(n)}$ are constants with symmetric indices $\bfa(n)$, is  by the property \eqref{lan} equivalent to the classical function
\be \label{AA*}
f(\bfcL) = [f_\* (\bfcL)]_W \,.
\ee
The temporal evolution of observables in phase space is generated by the the Hamiltonian $H$ by means the equivalent of the Heisenberg equation,
\be \label{evol}
\dot{f}=\frac{\partial f}{\partial t}  - \frac{ [H ,\, f]_\*}{i \ssl}\, . \qquad \dot{f}:= \frac{df}{dt}
\ee
In the classical limit $\ssl \rightarrow 0$ the latter is reduced to the Hamilton equation.

The solution of \eqref{evol} is
\be 
f(t,q,p)= \exp_\*(tH/i\ssl) \* f(0,q,p) \* \exp_\*(-tH/i\ssl) \,,
\ee
for the boundary value $f(0,q,p)$.

The Moyal equation, 
\be\label{Moyal}
\frac{\partial W}{\partial t} - \frac{[H,\,W]_\*}{i\ssl}= 0\,,
\ee
implies that the Wigner distribution is a constant of motion, and its classical limit reproduces the Liouville equation.

The static Wigner function satisfies $\partial W / \partial  t  =0$ and hence,
\be\label{staticM}
[H,\,W]_\*= 0\,,
\ee
which is solved also by solutions  of the  $\*$-genvalue equation \cite{Fairlie:1964qt},
\be \label{Wignereq}
H \* W =W\* H =  E \, W \, ,
\ee
which provides the spectrum of the Hamiltonian. \eqref{Wignereq} is, after a explicit computation of the $\*$-product, equivalent to the eigenvalue problem of a differential operator.

The following lemmas may be useful for the computations performed in this paper:

\begin{itemize}

\item { the multiplication of a function by the square of phase space coordinates \eqref{Ln} yields:}

\be \label{LL*f}
\bfcL_\bfa \bfcL_\bfb \* f(\cL) = \bfcL_\bfa \bfcL_\bfb  f(\cL) + i {\ssl}  \bfcL_{(\bfa}  \frac{\partial }{\partial \bfcL^{\bfb)}} f(\bfcL) -\frac{{\ssl}^2}{4} \frac{\partial ^2 }{\partial \bfcL^\bfa \partial \bfcL^\bfb} f(\bfcL) \,,\qquad \, 
\ee
\be \label{g*LL}
 f(\cL)\* \bfcL_\bfa \bfcL_\bfb = \bfcL_\bfa \bfcL_\bfb  f(\cL) - i {\ssl}  \bfcL_{(\bfa}  \frac{\partial }{\partial \bfcL^{\bfb)}} f(\bfcL) -\frac{{\ssl}^2}{4} \frac{\partial ^2 }{\partial \bfcL^\bfa \partial \bfcL^\bfb} f(\bfcL) \,,\qquad \, 
\ee

\item the fourth order  $\*$-product of a symplectic vector: 

\begin{eqnarray} \label{LL*LL}
\bfcL_\bfa \bfcL_\bfb \* \bfcL_\bfxi \bfcL_\bfz & =& \bfcL_\bfa \bfcL_\bfb  \bfcL_\bfxi \bfcL_\bfz +i \frac{{\ssl}}{2}\left(
\bfC_{\bfa \bfxi} \bfcL_\bfb \bfcL_\bfz + \bfC_{\bfa \bfz} \bfcL_\bfb \bfcL_\bfxi + \bfC_{\bfb \bfxi} \bfcL_\bfa \bfcL_\bfz + \bfC_{\bfb \bfz} \bfcL_\bfa \bfcL_\bfxi 
\right) \nonumber \\[4pt]
&& - \frac{{\ssl}^2}{4}\left(
\bfC_{\bfa \bfxi} \bfC_{\bfb \bfz}+\bfC_{\bfa \bfz} \bfC_{\bfb \bfxi}   \right) \, .
\end{eqnarray}

\end{itemize}
where we made use of the relations
$\frac{\partial \bfcL_\bfb}{\partial \bfcL^\bfa} = \bfC_{\bfa \bfb}  $ and 
 $\bfC_{\bfa \bfb} \frac{\partial f(\bfcL)}{\partial \bfcL_\bfb} = \frac{\partial f(\bfcL)}{\partial \bfcL^\bfa}$.

\subsection*{Example: Quantization of the harmonic oscillator}

Here we will review quantization of the harmonic oscillator (see \cite{Bayen:1977hb,Curtright:2011vw}).
The harmonic oscillator in  $d$ spatial dimensions has Hamiltonian,
\be\label{HHO}
H_d:=\frac{1}{2} \delta^{\bfa \bfb} \bfcL_\bfa \, \bfcL_\bfb  =\frac{1}{2}(\ap^2+\aq^2)\,, \qquad \aq \, , \ap \; \in \mathbb{R}^d\,. 
\ee
Here we introduced a standard notation for the $SO(d)$ invariant product in  $\mathbb{R}^d $,
\be 
\av^2 := \av \cd \av= (v_1)^2+ \cdots + (v_d)^2\,,  \qquad \av=(v_1, \cdots, v_d) \, \in \, \mathbb{R}^d \,.
\ee
  
With the help of \eqref{LL*f}, we find the $\*$-eigenfunctions of $H$,
\be \label{HOev}
H_d \* W_{n_1,\cdots , n_d} (\aq,\ap) = E_{n_1,\cdots , n_d} \,  W_{n_1,\cdots , n_d} (\aq,\ap)  \, ,\qquad E_{n_1,\cdots , n_d}= {\ssl}\, \Big({n_1+\cdots + n_d} +\frac{d}{2}\Big)\,. 
\ee
where 
\be \label{Wn1d}
 W_{n_1,\cdots , n_d} (\aq,\ap) := \Pi_{k=1}^d  \; W_{n_k}(q_k,p_k)\,,
\ee
is the classical product of solutions of the $d=1$-dimensional problem,
\be \label{Wn}
W_n(q,p)= \frac{(-1)^n}{\pi {\ssl}} \exp \Big(-\frac{p^2+q^2}{{\ssl}} \Big)  \, L_n\Big( \frac{p^2+q^2}{{\ssl}/2}\Big)\,, \qquad  L_n(z):=  \frac{1}{n!}\, \exp (z)\, \frac{d^n\:}{dz^n}(z^n \exp (-z))\,,
\ee
being $L_n$  polynomials of Laguerre. They satisfy the completion identity,
\be \label{compW}
\sum_{n=0}^\infty W_n= {\ssl}^{-1} \,.
\ee
Note that though the resulting spectrum of energies is the same as in canonical quantization, the $\*$-genfunctions of the Hamiltonian involve Laguerre and not Hermite Polynomials. The  relation between both kinds of polynomials  is given by \eqref{Wpsi}.

\section{Twistor--phase space (TPS)}\label{TPS}

Points on light-cones embeddings  in $2+1$ and $3+1$ space-time dimensions can be parametrized by Penrose's \cite{Penrose:1967wn,Penrose:1972ia} twistor space. The  coordinates of twistor space are also equivalent to coordinates of phase spaces (up to reality conditions). Indeed, two- and four-dimensional phase spaces have the same number of components than spinors in the respective dimensions and they transform accordingly under the respective (isomorphic) Lorentz algebras, $so(2,1) \cong sp(2)$ and $so(3,1) \subset so(3,2)  \cong sp(4)$.  Here we extend this twistor--phase space correspondence to higher dimensional phase spaces, which therefore will serve as moduli space for the construction of  higher dimensional  algebraic geometries beyond light-cones. 
To the the phase space regarded as the underlying space of these new algebraic geometries we shall refer as to  \textit{twistor--phase space} (TPS). This means, phase space vectors will be regarded as commuting spinors,  the symplectic matrix in phase space will be used as the conjugation matrix, and with the help of the elements of the exterior algebra of Dirac matrices we will construct maps from square products of phase space variables to sets of multivectors in Minkowski space, to be regarded as defining the coordinates of the algebraic geometry to be denoted $\bX[\mathcal{M}_D]$. There will be quadric identities satisfied by the multivectors as consequence of Fierz identities the exterior algebra of $\Gamma$-matrices. These identities will  characterize the space $\bX[\mathcal{M}_D]$.

We start studying the lower dimensional cases, $D=2+1$ and $D=3+1$ dimensions, on which the spaces $\bX[\mathcal{M}_D]$ consists of the light-cone and a subspace of the Grassmannian $\texttt{Gr}(2,4)$.  Then we generalize these results to dimensions $D=2,3,4 \texttt{ Mod } 8$, in correspondence with the existence of Majorana representations of Clifford algebras.

\subsection{The light-cone in $2+1$ dimensions}

The light-cone in $2+1$ dimensions is defined by the algebraic constraint, 
\be \label{cone3}
x^\mu x_\mu = 0 \,, \qquad \mu=0,1,2,
\ee
where the indices of the coordinates $x^\mu$ are contracted using the Minkowski metric $\eta_{\mu\nu}$, with diagonal elements $(-1,1,1)$.

 \eqref{cone3} can be solved parametrizing the two-dimensional light-cone using quadratic polynomials in phase space, i.e. 
\begin{equation}\label{xmuqp}
x^\mu=\frac{1}{4} \Big(\, p^2+q^2\,, \;p ^2-q^2 \,, \; -2 qp \, \Big)  \,.
\end{equation}
The coordinate $x^0$, the time, takes only positive values, hence \eqref{xmuqp} parametrizes the upper sheet of the light-cone. The map \eqref{xmuqp} covers the cone twice, which is reflected in the symmetry $(q,p)\rightarrow (-q,-p)$ as we can observe when we pass to polar coordinates,
\be \label{polar}
p= r \cos(\theta)\,, \qquad q=r \sin(\theta)  \, , 
\ee 
so
\begin{equation}\label{xmupolar}
x^\mu=\frac{1}{4} r^2  \Big(\, 1 \,,\cos(2\theta)\,, - \sin(2\theta) \Big)  \,. 
\end{equation}
Hence the phase space is a double cover of the (positive-time) light-cone. 
Phase space vectors will transform in a spin-$1/2$ representation of the Lorentz algebra, and therefore their symmetric quadratic products \eqref{xmuqp} will transform in a spin-one representation. This is more evident when we express  \eqref{xmuqp} in the form,  
\begin{equation}\label{xmu}
x^\mu=\frac{1}{4} \bar{{\cal L}}\, \gamma^\mu {\cal
L}\, , 	
\end{equation}
{ where $\bar{{\cal L}}^\a:=C^{\a\b}{\cal L}_\b$ is obtained from the phase space coordinates \eqref{Ln} and the convention  \eqref{sM2} to rise and lower indices, and where the} $2+1D$ Dirac matrices are in the Majorana representation
$$
(\gamma_0)_{\alpha}^{\ \beta}= -i(\sigma^2)_{\alpha}^{\ \beta},
\qquad (\gamma_1)_{\alpha}^{\ \beta}= (\sigma^1)_{\alpha}^{\
\beta}, \qquad (\gamma^2)_{\alpha}^{\ \beta}=-
(\sigma^3)_{\alpha}^{\ \beta}.
$$
 They satisfy
$$(\gamma_\mu)_\alpha{}^\xi(\gamma_\nu)_\xi{}^\beta=
    \eta_{\mu\nu}\delta_{\alpha}{}^\beta+\epsilon_{\mu\nu\lambda}
    (\gamma^\lambda)_\alpha{}^\beta\, ,$$
where for the Levy-Civita pseudo-tensor $\epsilon_{012}=1$.  

Now ${\cal L}$  looks like a twistor and the symplectic matrix $C_{\a\b}$ appears as the conjugation matrix. We rise and lower indices in North-West--South-East convention (the same than in \eqref{sM2}): $(\gamma_\mu)^{\beta \alpha}= C^{ \a \xi } (\gamma_\mu)_\xi{}^\b$, or $(\gamma_\mu)_\alpha{}^\xi C_{\xi \b}$, obtaining symmetric real matrices,  $ (\gamma_\mu)^{\alpha\beta}=(\gamma_\mu)^{\beta \alpha} $, $(\gamma_\mu)_{\alpha\beta}=(\gamma_\mu)_{\beta \alpha}$, $\tau^*=\tau_\mu$. 

That \eqref{xmu} solves \eqref{cone3} can be verified also using the Fierz identity,
\be
(\gamma^\mu)_\alpha{}^\beta (\gamma_\mu)_{\alpha '}{}^{\beta '}=2 \delta_\alpha{}^{\beta'} \; \delta_{\alpha'}{}^\beta - \delta_\alpha{}^{\beta} \; \delta_{\alpha'}{}^{\beta'},
\ee
 and that ${\cal L}_\alpha {\cal L}^\alpha =0$.

The $Sp(2)$ transformations in phase space,
\be \label{sp2L}
 \cL'_a= g_\a{}^\b \cL_\b \, , \qquad g:= \exp (\varepsilon^\mu J_\mu) \,,
 \ee
is generated by the matrices
\be
 J_\mu:=\frac{\gamma_\mu}{2}\,,
 \ee ~
which satisfy the $so(2,1) \cong sp(2)$ algebra,
\be\label{so12cliff}
[J_\mu,\, J_\nu]= \epsilon_{\mu\nu\lambda} J^\lambda \,,
\ee
and with parameters $\varepsilon^\mu$. 
\eqref{sp2L} induces the transformation,
\begin{equation}\label{xmu'}
x'^\mu=\frac{1}{4} {\cal L}'^\alpha (\tau^\mu)_\alpha{}^\beta {\cal
L}'_\beta \, = \Lambda^\mu{}_\nu x^\nu \, ,  
\end{equation}
which is equivalent to
\be
\Lambda^\mu{}_\nu= (\exp (\varepsilon^\lambda  \mathcal{J}_\lambda ) )^\mu{}_\nu \, , 
\ee
i.e. the exponential of  the spin-$1$ Lorentz algebra generated by the Levy-Civita pseudotensor,
\be
 (\mathcal{J}_\lambda)^{\mu}{}_\nu:= \e^\mu{}_{\lambda \nu} \,. 
\ee

\subsection{The light-cone in $3+1$ dimensions and the Grassmannian $\texttt{Gr}(2,4)$}
 
To be systematic, consider now a four dimensional twistor phase space, spanned by vectors { (cf. \eqref{Ln})}
\be \label{L2}
\bfcL_{\bfa}=(q_1, q_2,p_1,p_2)\, \quad \in \quad \mathbb{R}^{4}\,, 
\ee
and proceed as before.
The embedding of the light-cone in four dimensions is now given by,
\begin{eqnarray}\label{X}
 X^\bmu= \frac{1}{4} \bar{\bL}\gamma^\bmu\bL \,, \qquad \ba=1,2,3,4\,, \quad \bmu=0,1,2,3,
\end{eqnarray}
where 
$$
{\small
(\gamma^0)_\bfa{}^\bfb=\left(%
\begin{array}{cc}
  0 & \sigma^0 \\
  -\sigma^0 & 0 \\
\end{array}%
\right),\qquad
(\gamma^1)_\bfa{}^\bfb=\left(%
\begin{array}{cc}
  0 & \sigma^0 \\
  \sigma^0 & 0 \\
\end{array}%
\right)},$$
$$
{\small
(\gamma^2)_\bfa{}^\bfb=\left(%
\begin{array}{cc}
  \sigma^3 & 0 \\
  0 & -\sigma^3 \\
\end{array}\right),\qquad
(\gamma^3)_\bfa{}^\bfb=\left(%
\begin{array}{cc}
  -\sigma^1 & 0 \\
  0 & \sigma^1 \\
\end{array}%
\right)}.$$ 
are Dirac matrices in a  Majorana representation. They have the symmetry,
\be \label{Cgamma}
(\gamma^\bmu)^{\bfa \bfb}:=\bfC^{\bfa \bxi}(\gamma^\bmu)_{\bxi}{}^\bfb\,  \qquad (\gamma^\bmu)_{\bfa \bfb}:=(\gamma^\bmu)_{\bfa}{}^{\bxi} \bfC_{\bxi \bfb} \, , \quad (\gamma^\bmu)^{\bfa \bfb}=(\gamma^\bmu)^{\bfb \bfa},\quad (\gamma^\bmu)_{\bfa \bfb}=
(\gamma^\bmu)_{\bfb \bfa},
\ee
and are real-valued. Again, we have used the symplectic matrix \eqref{sM1} as the conjugation matrix of the Clifford algebra generated by $\gamma^\bmu$'s, which is possible owing to the algebra isomorphism  $sp(4) \cong so(3,2)$. 

From \eqref{X} we obtain,
\begin{eqnarray}\label{XXXX}
&X^{0}=\frac{1}{4}(p_1^2+q_1^2+p_2^2+q_2^2),\qquad
X^{1}=\frac{1}{4}(p_1^2-q_1^2+p_2^2-q_2^2),&\nonumber
\\[6pt]
&X^{2}=\frac{1}{2}(q_1p_1-q_2 p_2),\qquad
X^{3}=-\frac{1}{2}(q_1p_2+q_2p_1)\,.&
\end{eqnarray}
It  can be verified that 
\begin{eqnarray}\label{XX}
 X^\bmu X_\bmu =0\,.
\end{eqnarray}

The light-cone in four dimensions is a three dimensional manifold, the ``world volume" of a sphere expanding at the speed of light. Since the phase space behind its construction  is four dimensional one  degree of freedom is missed by the map \eqref{XXXX}. Indeed, it is contained in the remaining second order polynomials  
\begin{eqnarray}\label{Z}
 Z^{\bmu\bnu}=\frac{1}{4} \bar{\bL} \gamma^{\bmu\bnu} \bL \,,\qquad \gamma^{\bmu\bnu} := \frac{1}{2} [\, \gamma^\bmu\,, \, \gamma^\bnu \,] \,.
\end{eqnarray}
Only gamma matrices and their commutators yield non-trivial result when the double contractions \eqref{X}-\eqref{Z} are performed.

By their construction,
\begin{eqnarray}
Z^{0i}&=&\left( \,-\frac{1}{2} (q_1 p_1  + q_2 p_2), \, \frac{1}{4} ( p_1^2 - q_1^2  - p_2^2+ q_2^2), \,
 \frac{1}{2}(  q_1 q_2 -p_1 p_2)\right) \label{Z0i} \\
Z^{i}&=&\left( \frac{1}{2}  (q_1  p_2 -  q_2 p_1),\, -\frac{1}{2}  (p_1 p_2 + q_1 q_2), \,
 \frac{1}{4}  (-p_1^2- q_1^2  + p_2^2 + q_2^2 ) \right) \, , \label{Zi}
\end{eqnarray}
where we have defined
\be \label{Zi}
Z^{i}:=\tfrac{1}{2}\epsilon_{ijk} Z^{ij} = (Z^{23}, Z^{31}, Z^{12} )\, , \quad i,j,k=1,2,3,
\ee
and $\epsilon_{ijk}$ is the euclidean Levy-Civita pseudo-tensor. 

The tensor $Z^{\bmu\bnu}$ has four non-equivalent degrees of freedom. The redundancy of its six components is reflected in the identities $Z_{01} Z_{ 23}+Z_{12} Z_{03} + Z_{20} Z_{ 13}  =0 \,,$ equivalent to 
\be \label{Pluck2}
Z^{[\bmu \bnu} Z^{ \blambda \brho]} =0 \,,
\ee
and 
\be \label{ZZ}
 Z^{\bmu \bnu} Z_{ \bmu \bnu} =0 \,,
\ee
where the Levi-Civita pseudotensor is such that $\epsilon_{0123}=1$. 
 The identity \eqref{ZZ} is consequence of the identities 
 \be 
X^i X_i = Z^{0i} Z^0{}_i=(X^0)^2 \, ,\qquad  X^i Z^0{}_i= 0 \,.
\ee 
The constraints of the form \eqref{Pluck2} are  known as the  Pl\"ucker constraints, and the respective bivector coordinates describe planes passing through the origin of a four dimensional space. The space of all these planes form the Grassmannian $\texttt{Gr}(2,4)$ variety\footnote{The generalization of this variety to higher dimensional $k$-planes through the origin of a $n$-dimensional vector spaces was studied by H. Grassmann and denoted Grassmannian $\texttt{Gr}(k,n)$.}.
 
The bivector \eqref{Z} describe in reality a subspace of $\texttt{Gr}(2,4)$ since the identity, also satisfied from definitions \eqref{XXXX} and \eqref{Z},
\be \label{epsiXZ}
 X^{[\bmu } \, Z^{\bnu \blambda]} =0 \,,
\ee
means that the planes $Z^{\bmu \bnu}$ contain at least one light-ray. However, in $\texttt{Gr}(2,4)$ there are also other planes which do no contain light-rays. To the subspace of $\texttt{Gr}(2,4)$ of planes containing light-rays we refer as to $\texttt{Gr}_\texttt{LC}(2,4)$.

In more detail, Pl\"ucker coordinates are equivalent to the exterior product
\be \label{cZ}
Z^{\bmu \bnu}:=  (v^1)^\bmu \, (v^2)^\bnu -  (v^2)^\bmu \, (v^1)^\bnu \,,
\ee
of two non-parallel  defining vectors $v^1$ and $v^2$. To obtain \eqref{Z} we identify 
\be \label{v1v2}
(v^1)^\bmu= X^{\bmu }\,, \qquad  (v^2)^\bnu=\frac{1}{X^0}(0, Z^{01}\,,
  Z^{02}\,, Z^{03}),
\ee
which demonstrate that the light-ray $X^\bmu$ is in the plane  $Z^{\bmu \bnu}$.

\subsection{Phase space and the multivector-extended space-time}\label{sec:XVM}

Similarly to the coordinates of the light-cone in $D=3,\,4$ and of the bivector \eqref{Z}, the Dirac matrices permit more generally the construction of mappings from $2^{[D/2]}$-dimensional phase space to a geometry formed by multivector in $D$-Minkoswki space-time. Here we shall construct the correspondent multivector coordinates and identify the quadratic identities that they satisfy.

The multivector coordinates appear from the contraction of polynomials of second order in phase space and the exterior algebra of Dirac-gamma matrices,
\be \label{Gamman}
\Gamma^{\bmu[n]}:=\Gamma^{[\bmu_1} \cdots \Gamma^{\bmu_n]}= \frac{1}{n!} \sum_{P\{n\}} (-1)^P\,  \Gamma^{P(\bmu_1)} \cdots \Gamma^{P(\bmu_n)}\,,
\ee
where $P( \cdot )$ denotes an element of the permutation group $P\{n\}$ of $n$ elements and $P=0,1$ its parity.  We assume the convention $\{ \Gamma_\bmu \,, \Gamma_\bnu \}= 2\ \bss{\eta}_{\bmu \bnu}$ with metric diagonals entries $ (-1,1,\cdots ,1).$ 
The multivector coordinates are given by the double contraction of phase space vectors \eqref{Ln} and the matrix indices {  of $(\Gamma^{\bmu[n]})_\bfa{}^\bfb$},
\be \label{Xmun}
\bX^{\bmu[n]} := \frac{1}{4} \bar{\bL}\, \Gamma^{\bmu[n]}  \bL \,,\qquad n=1,2 ~ \texttt{Mod} ~ 4 \,, \quad   D=2,3,4\, ~ \texttt{Mod} ~ 8 \,.
\ee 
The reality condition implies that the gamma matrices should be in the Majorana representation which implies the correspondent restrictions on the dimension of the space-time and the rank $n$ of the multi-vectors (see \cite{vanHolten:1982mx}). Indeed only for $n = 1,2 ~ \texttt{Mod} ~ 4$ \eqref{Xmun}  yields a non-trivial result. This can be seen by rising one of the spinor indices, using the conjugation matrix $\bfC$ defined in \eqref{sM1}. According to the convention \eqref{sM2} they satisfy
\be \label{CGammaC}
(\Gamma^{\bmu[n]})^{\ba \bb} =(-1)^{[(n-1)/2]} (\Gamma^{\bmu[n]})^{\bb \ba} \,,
\ee
{ In particular we chose a representation in which $(\Gamma^{0})^{\ba \bb}=\delta^{\ba\bb}$, so that $\bX^0\geq 0$, since,
$$
\bX^0= (\ap^2+\aq^2)/4\,,
$$
is proportional to the energy of a harmonic oscillator (see also \eqref{HHO}). Therefore, the coordinates $\bX^\bmu$ correspond to points on the non-negative--time half hyper-plane of the $D$-Minkoswki space.  The chiral Gamma matrix is defined as,
$$
\widetilde{\Gamma}= \Gamma^0\Gamma^1\cdots \Gamma^{D-1},
$$
which upon $\Gamma^{\bmu[n]}$ produced the Hodge duals,
\be \nonumber
\Gamma^{\bmu[n]}\widetilde{\Gamma}=(-1)^{n(n+1)/2} \star \Gamma^{\bmu[n]}\, ,\qquad 
\star \Gamma_{\bmu[n]} :=\frac{1}{(D-n)!}\epsilon_{\bmu[n]}{}_{\bnu[D-n]}  \Gamma^{\bnu[D-n]}\,.
\ee
The equation \eqref{CGammaC}} implies $sp(2^{[D/2]})$-algebra reality conditions for the symmetric case,
  \be \label{Gammasp}
 \bfC \Gamma^{\bmu[n]} +  (\Gamma^{\bmu[n]} )^t \bfC = 0\,,\qquad n=1,2 ~ \texttt{Mod} ~ 4\,.
\ee
Thus there are $\texttt{dim} ~ (sp(2^{[D/2]})$ non-trivial multivector coordinates $\bX^{\bmu[n]}$. The space  parametrized by multivector coordinates,
\be \label{XMD}
 \bX [\mathcal{M}_D]=\left\{\bX^{\bmu[n]}=\frac{1}{4} \bar{\bL}\Gamma^{\bmu[n]}  \bL \, , \quad  n = 1,2 ~ \texttt{Mod} ~ 4 \, \leq \, n_{\texttt{max}}, \quad   D=2,3,4\, ~ \texttt{Mod} ~ 8 \right\}\,, 
 \ee
has, by construction, dimension
\be
 \texttt{dim} ~ ( \bX[\mathcal{M}_D])=2^{[D/2]}\, ,
\ee 
since  $\bX[\mathcal{M}_D]$ has $2^{[D/2]}$ phase space $\bL$-parameters.  Hence there must be $\texttt{dim} ~ (sp(2^{[D/2]}) - 2^{[D/2]}$  identities to be satisfied by the multivectors \eqref{XMD}.

Let us define the unconstrained multivector space
\be \label{VD}
\bV[\mathcal{M}_D]=\bigoplus^{n_{\texttt{max}}}_{n=1,2 ~ \texttt{Mod} ~ 4} V^{[n]}\,,
\ee
which is formed by the direct sum of spaces $V^{[n]}$ of unbounded multivectors $Y^{\bmu[n]}$ of rank $n$. Here $n_{\texttt{max}}$ is the \textit{supremum} of integers $n$ of type $1,2 ~ \texttt{Mod} ~ 4 $ such that $n \leq D$ for $D$ even or  $n\leq [D/2]$ for  $D$ odd.  Thus $\bV [\mathcal{M}_D]$ is an extension of space-time with multivector extra dimensions and $\bX [\mathcal{M}_D] \subset \bV [\mathcal{M}_D]$. 

$\bV[\mathcal{M}_D]$ has dimension
\be \label{card}
 \texttt{dim}~ ( \bV[\mathcal{M}_D])  =\texttt{dim}~ (sp(2^{[D/2]})) = 2^{[D/2]-1} (2^{[D/2]}+1) \,.
\ee 
The 
\be \label{dimV-X}
 2^{[D/2]-1} (2^{[D/2]}-1)=\texttt{dim}~ ( \bV[\mathcal{M}_D]) - \texttt{dim}~ (\bX[\mathcal{M}_D] )\,
\ee
algebraic identities satisfied by  multivectors in $ \bX [\mathcal{M}_D]$ can be regarded as restrictions on the space which define it as a subspace of $\bV[\mathcal{M}_D]$, i.e.
\be \label{Vrestriction}
 \bX [\mathcal{M}_D] = \{  \bX^{\bmu[n]} \, \in \, \bV [\mathcal{M}_D] \: | \: \{\bf \Lambda \}( \bX^{\bmu[n]} )=0 \, \}\,,
\ee
where $\{\bf \Lambda \}$ represents the $2^{[D/2]-1} (2^{[D/2]}-1)$ quadratic algebraic restrictions to be found. The TPS parametrization \eqref{XMD} should be regarded as  an efficient way to solve the system of constraints, analogously to twistor space with respect to light-cones. These identities will be derived from Fierz identities in the space of matrices $\Gamma^{\bmu [n]}$ and the symmetry of the classical product of phase space vectors.

Using the matrices $\Gamma$ we can decompose the square products of phase space vectors in terms of Minkowski space multivectors as
\be \label{LaLb}
\bL_\ba \bL_\bb = \frac{1}{2} \{ \bL_\ba, \bL_\bb\}_\*  = \frac{4}{2^{[D/2]}} \sum _{n=1,2 ~ \texttt{Mod} 4}' \frac{(-1)^{[n/2]}}{n!}   (\Gamma^{\bmu[n]})_{\ba\bb} \, \bX_{\bmu[n]} \,,
\ee
where prime-$'$ in the sum means as in \cite{vanHolten:1982mx}, that it should restricted to integers $n \leq D$ for $D$ even, and  $n\leq [D/2]$ for $D$ odd. \eqref{LaLb} can be verified contracting on both sides of \eqref{LaLb}  with $(\Gamma^{\bmu[n]})^{\ba \bb}$, which yields the definition of $\bX^{\bmu[n]}$ on the l.h.s., and on the r.h.s. it yields traces of products of gamma matrices, i.e.
\be \label{tr} 
\texttt{Tr}\ (\Gamma^{\bmu[n]}\Gamma_{\bnu[m]} )= (-1)^{[n/2]} 2^{[D/2]} \delta^n_m\, \delta^{\bmu[n]}_{\bnu[n]}\,.
\ee 
Here $\delta^{\bmu[n]}_{\bnu[n]}=\delta^{\bmu_1\cdots \bmu_n}_{\bnu_1\cdots \bnu_n}$ is the Kronecker symbol. 

Since $\bL$ in \eqref{LaLb}'s are continuous parameters we can take two derivatives to obtain the identity
\be \label{Ide1} 
\bfC_{\ba \bb} \bfC_{\ba' \bb'} +\bfC_{\ba' \bb} \bfC_{\ba \bb'} =\frac{1}{2^{[D/2]-1}} \sum _{m=1,2\,\, ~ \texttt{Mod} ~ 4}' \frac{(-1)^{[m/2]+1}}{m!} (\Gamma^{\bnu[m]})_{\ba \ba'}  (\Gamma_{\bnu[m]})_{\bb \bb'}\ ,
\ee
and from the latter, multiplying by gamma matrices and appropriated contraction of spinor indices, we obtain;
\bea \label{Ide2} 
&&(\Gamma^{\bmu[n]})_\ba{}^{\bb '}(\Gamma_{\bmu[n]})_{\ba'}{}^{\bb } +(\Gamma^{\bmu[n]})_{\ba \ba ' }(\Gamma_{\bmu[n]})^{\bb \bb '} = \nonumber \\[4pt]
&& \hspace{3cm} \sum _{m=1,2\,\, ~ \texttt{Mod} ~ 4}'  \frac{(-1)^{[(m+n)/2]}}{2^{[D/2]-1}}   \frac{n!}{m!} \, h(D,n,m)\,  (\Gamma^{\bnu[m]})_{\ba '}{}^{ \bb '}  (\Gamma_{\bnu[m]})_{\ba }{}^{ \bb } \,.
\eea
Here the coefficient
\be
h(D,n,m):= \sum_l (-1)^l \binom{m}{l} \binom{D-m}{n-l}\,,
\ee
where $l$ takes values in the rank of integers that does not yield ill-defined binomial coefficients, can be  obtanined using the constant of structure of the associative algebra of gamma matrices \ref{app:Gamma}  and from the identity (see e.g. \cite{vanHolten:1982mx})
\be \label{gngmgn}
\Gamma^{\bmu[n]}\Gamma^{\bnu[m]}\Gamma_{\bmu[n]}= n! (-1)^{nm+[n/2]}\, h(D,n,m)\, \Gamma^{\bnu[m]}\,.
\ee

Contraction of \eqref{LaLb} with the product $\bL^\ba \bL^\bb$, and from the identity $\bL^\ba \bL_\ba=0$, implies that
\be \label{hyper1}
\sum _{n=1,2\,\, \texttt{Mod} 4}' \frac{(-1)^{[n/2]}}{n!} \bX^{\bmu[n]}  \bX_{\bmu[n]} = 0\,.
\ee
This equation reproduces the light-cone constraint in $2+1$. In $3+1$ dimensions we obtain the sum of terms proportional to \eqref{XX} and \eqref{ZZ}, i.e. $X^\mu X_\mu=0$ and $Z^{\mu\nu} Z_{\mu \nu}=0$, which therefore solve the respective constraints. The latter case shows however that \eqref{hyper1} is not an independent constraint. Indeed, \eqref{hyper1} is just the simplest algebraic relation satisfied by multivectors $\bX^{\bmu[n]}$. In what follows we shall present a systematic way to identify a subset of independent linear combinations of squares  products of multivectors.

\subsection{The multivector quadric}

We can obtain a linear system of equations for the squares of multivector coordinates replacing $\bL_\ba  \bL^\bb$ in the product
\be
\bX^{\bmu[n]}  \bX_{\bmu[n]}=\frac{1}{16} \bL^{\ba'} (\Gamma^{\bmu[n]})_{\ba'}{}^\ba (\bL_\ba  \bL^\bb) (\Gamma_{\bmu[n]})^{\bb \bb'} \bL_{\bb'}\,,
\ee
by \eqref{LaLb}. 
Now using \eqref{gngmgn} we obtain
\be \label{X2MX2}
\bX^{\bmu[n]} \bX_{\bmu[n]} = \sum _{m=1,2\,\, \texttt{Mod} 4}' ({\bf M_D})_n{}^m\  \bX^{\bmu[m]}  \bX_{\bmu[m]}\,,
\ee
where 
\be \label{Mjn}
({\bf M_D})_n{}^m = \frac{(-1)^{[(m+n)/2]}}{2^{[D/2]}}\frac{n!}{m!}\, \, h(D,n,m) \,.
\ee
Let us define ${\bf M_D}$ as the matrix with the latter entries and subtract the identity matrix
\be\label{Lbda}
{\bf \Lambda_D := M_D} - \mathds{1}\,.
\ee
This together with the column matrix
\bea 
{\bf V_D}:= \left(\begin{array}{c}
\bX^{\bmu} \bX_{\bmu} \\[4pt]
\bX^{\bmu[2]} \bX_{\bmu[2]}\\
\vdots
\end{array}\right)\, ,
\eea
permits to express \eqref{X2MX2} as a homogeneous equation;
\be\label{Lbda-1}
{\bf \Lambda_D \ V_D}=0 \,.
\ee
Since ${\bf \ V_D}$ belongs to the kernel of the matrix ${\bf \Lambda_D }$, there should be $\texttt{rank}({\bf M_D})$ independent constraints on linear combinations of the Lorentz-scalars $\bX^{\bmu[n]}\bX_{\bmu[n]}$. This reasoning applies also to linear combinations of quadratic products  of multivectors of different orders. 

Let us multiply in both sides of the equality \eqref{Ide2} by the matrix {  ${\cal T}={\cal T}_\kappa{}^\ba:=\tau^{\bmu[n]}(\Gamma_{\bmu[n]})_\kappa{}^{\bfa}$, where $\tau^{\bmu[n]}$ are constant (multivector) parameters, then we contract the spinor index labeled by $\bfa$,
\bea \nonumber
&&({\cal T}\Gamma^{\bmu[n]})_\kappa{}^{\bb '}(\Gamma_{\bmu[n]})_{\ba'}{}^{\bb } +({\cal T}\Gamma^{\bmu[n]})_{\kappa \ba ' }(\Gamma_{\bmu[n]})^{\bb \bb '} = \nonumber \\[4pt]
&& \hspace{3cm} \sum _{m=1,2\,\, ~ \texttt{Mod} ~ 4}'  \frac{(-1)^{[(m+n)/2]}}{2^{[D/2]-1}}   \frac{n!}{m!} \, h(D,n,m)\,  (\Gamma^{\bnu[m]})_{\ba '}{}^{ \bb '}  ({\cal T}\Gamma_{\bnu[m]})_{\kappa }{}^{ \bb } \,,\nonumber
\eea
resulting is an expression that has four free spinor indices. Similarly, using the the chiral matrix, we define $\widetilde{{\cal T}}=\widetilde{{\cal T}}_\kappa{}^\ba:= \widetilde{\Gamma}{\cal T}$, and multiply \eqref{Ide2} by  $\widetilde{{\cal T}}$, contract the index $\bfa$. In both cases, the four free spinor indices  can be contracted with the product of four phase space coordinates $\bL^\kappa \bL^{\bfa'}\bL_{\bfb}\bL_{\bfb'}$,  which will induce, from identity \eqref{Ide2}, new identities among multivectors,
\be \label{XZ}
\bX^{\bmu[n]} \bZ^T_{\bmu[n]} = \sum _{m=1,2\,\, \texttt{Mod} 4}' ({\bf M_D})_n{}^m\  \bX^{\bmu[m]}  \bZ^T_{\bmu[m]}\,,
\ee
where 
\be \label{Znr}
\bZ^T_{\bmu[n]}:= \bar{\bL} \Gamma_{\bmu[n]} T  \bL \,,
\ee
 are defined analogously to \eqref{Xmun} but now in terms of the phase space coordinates \eqref{Ln} 
and the matrix $T$, which is either $T={\cal T}$ or ${T}=\widetilde{\cal T}$, for all parameters $\tau^{\bnu[n]}$.
Since in odd dimensions $\widetilde{\Gamma}$ is proportional to the identity, the constraints \eqref{XZ} yield  by ${\cal T}$ and $\widetilde{\cal T}$ are equivalent}. In any case the introduction of $\widetilde{\Gamma}$ allows us to summarize \eqref{XZ} in a single equation,
\be\label{LVDT}
{\bf \Lambda_D} ~ {\bf V_D}^T=0 \ , \qquad
{\bf V_D}^{T}:= \left(\begin{array}{c}
\bX^{\bmu} \bZ^T_{\bmu} \\[4pt]
\bX^{\bmu[2]} \bZ^T_{\bmu[2]}\\
\vdots
\end{array}\right)\,,
\ee
 which in particular contains \eqref{Lbda-1} for $T=\mathds{1}/4$.
Observe that the system of constraints \eqref{LVDT} are all encoded in the matrix $\bLam_{\bf D}$ \eqref{Lbda}, which therefore characterizes the space $\bX[\mathcal{M}_D]$, up to some  redundancies produced by degeneracy of the matrices ${\bf \Lambda_D}$ and
the linear dependence among the ${\bZ}^{T}$ in odd dimensions since from the absence of a non-trivial chiral gamma matrix. Finally, we can regard \eqref{LVDT} as the definition of $\{\bf \Lambda \}( \bX^{\bmu[n]} )=0$ in \eqref{Vrestriction}.

In summary, the system of constraints \eqref{LVDT} that restrict \eqref{VD} to \eqref{XMD} can be systematically constructed as kernels of the matrix  ${\bf \Lambda_D}$  in all space-time dimensions $D=2,3,4 ~ \texttt{Mod } 8$,  and they can solved by \eqref{Xmun} in terms of TPS variables. 
 
\subsubsection*{Examples: $D=2,3,4, 10,11,12$} 
 
In $1+1$ dimensions we obtain from \eqref{Mjn} and \eqref{Lbda-1} 
\be 
{\bf \Lambda_2 }=\left(\begin{array}{cc}
-1 & \frac{1}2 \\[4pt]
1 & - \frac{1}2 
\end{array}\right)\ , \qquad D=1+1\ .
\ee
Since its rank is $1$ only one independent set of equations is produced;
\be \label{2dMV}
\bX^\bmu \bZ^T_\bmu -\frac{1}{2} \bX^{\bmu[2]} \bZ^T_{\bmu[2]}=0 \,,   
\ee
for fixed $T$ in \eqref{Znr}. It can be verified that only for $T=\mathds{1}$ a non-trivial and independent constraint is produced, while for others $T$'s it either reproduces the case $T=\mathds{1}$ or it yields trivial relations.

In $2+1$ dimensions $\bf \Lambda_3$ is a scalar, thus
\be \label{3dMV}
{\bf \Lambda_3} \, \bX^\bmu \bZ^T_\bmu = 0 \,,   \qquad {\bf \Lambda_3}=-\frac{3}{2}\,.
\ee
Indeed, \eqref{2dMV} and \eqref{3dMV} are equivalent. More generally, for a fixed dimension of the phase space, the characteristic constraints of multivectors $\bX^{\bmu[n]}$ in dimension $D=2 \ \texttt{Mod } 8$ and in one dimension higher $D=3~\texttt{Mod}~8$ are equivalent. The later is consequence of the equivalence of the exterior algebra of gamma matrices in the respective dimensions. Thus the spaces  $\bX [\mathcal{M}_D] \cong  \bX [\mathcal{M}_{D+1}]$ for $D=2 \ \texttt{Mod } 8$ are equivalent up to Lorentz covariant labels in  the respective dimensions. 

In the $3+1$ dimensional case,
\be
 {\bf \Lambda_4 }=
\left(
\begin{array}{cc}
 -\frac{3}{2} & 0 \\
 0 & -\frac{3}{2} \\
\end{array}
\right)\,.
\ee
Since ${\bf \Lambda_4}$ has rank two \eqref{LVDT} the independent constraints are just,
\be \label{4dMV}
\bX^\bmu \bZ^T_\bmu = 0\,, \qquad \bX^{\bmu \bnu} \bZ^T_{\bmu \bnu} = 0\,,   
\ee
which for different choices of $T$ it yields the previous results: \eqref{XX}, \eqref{Pluck2}, \eqref{ZZ} and \eqref{epsiXZ}.

For the cases $D=9+1, 10+1, 11+1$ the respective matrices ${\bf \Lambda_D}$ can be found in the appendix \ref{app:Lambda}. The resulting independent constraint, i.e. after elimination of redundant rows of the matrix ${\bf \Lambda_D}$, can be chosen as:
\bea 
 D&=&{}9+1 \; :  \qquad \left\{ \begin{array}{l}
1296 \bX^{\bmu[1]}\bZ^T_{\bmu[1]}+192 \bX^{\bmu[2]}\bZ^T_{\bmu[2]}- \bX^{\bmu[5]}\bZ^T_{\bmu[5]}=0\ , \\[4pt]
 8640 \bX^{\bmu[1]}\bZ^T_{\bmu[1]} + 720 \bX^ {\bmu[2]}\bZ^T_{\bmu[2]}-\bX^{\bmu[6]}\bZ^T_{\bmu[6]}=0\ , \\[4pt]
 362880 \bX^{\bmu[1]}\bZ^T_{\bmu[1]}-\bX^{\bmu[9]}\bZ^T_{\bmu[9]}=0\ , \\[4pt]
2903040 \bX^{\bmu[1]}\bZ^T_{\bmu[1]}+362880 \bX^{\bmu[2]}\bZ^T_{\bmu[2]}-\bX^{\bmu[10]}\bZ^T_{\bmu[10]}=0\ , 
\end{array}\right.
 \, \\[6pt]
 D&=&10+1 \; : \qquad \left\{ \begin{array}{l}
10 \bX^{\bmu[1]}\bZ^T_{\bmu[1]}+ \bX^{\bmu[2]}\bZ^T_{\bmu[2]}=0\ ,\\  720 \bX^{\bmu[1]}\bZ^T_{\bmu[1]}+ \bX^{\bmu[5]}\bZ^T_{\bmu[5]}=0\ ,
\end{array} \right.
 \\[6pt]
 D &=& 11+1 \; : \qquad \left\{ \begin{array}{l}
720 \bX^{\bmu[1]}\bZ^T_{\bmu[1]} + \bX^{\bmu[5]}\bZ^T_{\bmu[5]}=0\ , \\[4pt]
 2880 \bX^{\bmu[1]}\bZ^T_{\bmu[1]}+ 720 \bX^{\bmu[2]}\bZ^T_{\bmu[2]} + \bX^{\bmu[6]}\bZ^T_{\bmu[6]}=0\ , \\[4pt] 1814400 \bX^{\bmu[1]}\bZ^T_{\bmu[1]}-\bX^{\bmu[9]}\bZ^T_{\bmu[9]}=0\ , \\[4pt]
 14515200 \bX^{\bmu[1]}\bZ^T_{\bmu[1]}+1814400 \bX^{\bmu[2]}\bZ^T_{\bmu[2]}-\bX^{\bmu[10]}\bZ^T_{\bmu[10]}=0 \ .
\end{array} \right.
\eea
{ In order to obtain scalar identities, i.e. on the norms $\bX^{\bmu[n]}\bZ^T_{\bmu[n]}=\bX^{\bmu[n]}\bX_{\bmu[n]}$, for different $n$'s, we should choose $T=\mathds{1}/4$. Note that in the $9+1$ dimensional case, because there are four independent constraints and six products $\bX^{\bmu[n]}\bX_{\bmu[n]}$, $n=1,2,5,6,9,10$, only two of them can be chosen to be independent, say $\bX^{\bmu}\bX_{\bmu}$ and  $\bX^{\bmu[2]}\bX_{\bmu[2]}$. Because  $\bX^0 \geq 0$, the event-coordinates $\bX^{\bmu}$ parametrize a half of the Minkowski space. Analogous  analysis follows for $D> 10$. Hence, the multivector varieties constructed here contain the non-negative time half of the $D$ -dimensional Minkowski space-time for $D\geq 9+1$.}
 
 \section{Quantization in TPS} \label{QinTPS}

The first example of non-commutative space-time was proposed by Snyder \cite{Snyder:1946qz,Lu:2011it}, as a quantization of the quadratic constraint  that defines the de Sitter space as an embedding in $5$ dimensions. Snyder constructed differential operators (acting on a Hilbert space) as representing the coordinates of the de Sitter space together with the rank-two tensors operators generating Lorentz transformations, altogether spanning a representation of the algebra $so(4,1)$. Here we shall proceed in a different way. On the one hand, we will treat higher rank multivectors as  in the same category than  (vector) coordinates, i.e. similarly to Pl\"ucker (planes) coordinates.  On the other hand, we shall use TPS as a moduli of the multivector space already in the classical level, and then applying the phase space formulation of quantum mechanics presented in section \ref{PSQ} we shall quantize them. Thus non-commutativity will enter, not by means of an operator-correspondence but, by means of the $\*$-product. This is convenient regarding the classical limit since the $\*$-product becomes the standard commutative product of numbers and while multivectors will be kept invariant. 
 
Expressing the coordinates of the multivector geometries \eqref{XMD} in terms of TPS variables facilitate their quantization. First of all, the $\*$-product in TPS provides Lorentz covariant constants of structure to the associative algebra of multivectors. Without the TPS the task of constructing a suitable $\*$-product convolution-like covariant formula (cf. \eqref{f*g2n}) defined in terms of multivectors by themselves is a difficult task. Secondly, the $\*$-genvalue equation (solved by the Wigner function) provides natural prescription for the measurement of the expectation values of multivector coordinates and their functionals.
Indeed,  the ``star-genvalue" equation is given by
 \be \label{int}
G \* W = \textit{g}\, W\,,
\ee
 where  $g$ is the $\*${-genvalue} of the geometrical observable $G$. Here $G$ represents a function of TPS variables, or of multivectors, and $g$ is a number belonging to its spectrum. If  a functions $f$ is such that $[f,G]_\*=0$, then $W$ its also its $\*${-genfunction}. To mention an example, to measure the time we should use $G=\bX^0$ in the equation \eqref{int}. { Since $\bX^0$ is, as a function of phase space coordinates,  proportional to the $d$-dimensional Harmonic oscillator \eqref{HHO}, i.e. $\bX^{0}=H_d /2$,
solving the problem \eqref{int} yields,}
\be \label{X0spec}
\bX^{0} \* W_{n_1\cdots n_d}=\frac{1}{4} \ssl \Big(n_1+\cdots+ n_d + d \Big) W_{n_1\cdots n_d}\,, \qquad d=2^{[D/2]}\,.
\ee
Thus the Wigner functions encodes the perspective of a ``time-observer" on non-commutative space-like foliations, at constant discrete values of time, of spacetime.

The expectation (average) values of any observable in TPS, say $f$, will be measured with respect to the Wigner function by integrating
$$
\langle f \rangle_W = \int_{\texttt{TPS}} \, f \* W \,, \qquad \langle 1 \rangle_W=1\,.
$$
The variance of $f$ is then defined as
\be \label{Varf}
\texttt{Var}_{W}(f):= \langle f\* f \rangle_W - \langle f  \rangle_W ^2 \,.
\ee
For $G$ solving \eqref{int} and any $f$ such that $[f,G]_\*=0$, then their variance \eqref{Varf} vanishes. Hence the Wigner function provides the spectrum of the operators which are diagonal with respect to it and it minimizes their variance. { This can be expressed as a variational principle, }
\be \label{deltaVar}
\delta_W \texttt{Var}_W (G) =0\,, \quad \texttt{for} \quad G \* W = \textit{g}\, W\, .
\ee
 In this sense $\texttt{Var}_{W}(G)$ can be regarded as an ``action functional" for the Wigner function, and the operator $G$ as providing its ``kinetic" term. Note that the fact that due to the presence of the term $\langle G  \rangle_W ^2$ this principle is bi-local.
 
  In what follows we omit the label $W$  in $\langle \cdot \rangle_W$. 

\subsection{Quantization of the light-cone in $2+1$ dimensions}

From the definition of $x^\mu$ in terms of TPS \eqref{xmu} and the lemma \eqref{LL*LL}, we obtain 
the following deformation of the product of light-cone coordinates,
\be \label{asxx}
x^\mu\* \, x^\nu  = x^\mu\, x^\nu +  \frac{i{\ssl}}{2} \e^{\mu \nu}{}_\lambda  \,x^\lambda +\frac{{\ssl}^2}{16}  \eta^{\mu \nu} \,,
\ee
which involves only Lorentz-covariant structures, the coordinates, the Levy-Civita pseudo-tensor and the metric tensor. Note that, though non-perturbative, the $\*$-product provides corrections to the classical product $x^\mu\, x^\nu$ up to order ${\ssl}^2$. 

Now, the quantum deformation of the light-cone constraint \eqref{cone3} yields,
\be \label{xx3}
x^\mu \* x_\mu = \frac{3}{16}{\ssl}^2 \,,
\ee
equivalent to the one-sheet fuzzy hyperboloid, the non-compact counterpart of the \textit{fuzzy sphere} \cite{Hoppe1982,Madore:1991bw} (see also \cite{Bergshoeff:1989ns,Vasiliev:1989re}). \eqref{xx3} can be  regarded also as defining a two-dimensional \textit{fuzzy de Sitter} space \cite{Gazeau:2006hj,Jurman:2013ota} embedded in three dimensions. In any case, note that the conical singularity of the light-cone is smoothed out.

The anti-symmetric part of the $\*$-product \eqref{asxx} contains the commutation relation of the algebra $so(2,1)\cong sp(2) \cong sl(2, \mathbb{R})$,
\be \label{sl2}
[x^\mu\,, \, x^\nu]_\* = i{\ssl} \e^{\mu \nu}{}_{\lambda} \,x^\lambda \,.
\ee
Extending this algebra with the following relations,
\begin{eqnarray}\label{osp12a}
&
[x_{\mu},{\cal L}_{\alpha}]_{\* }=
-\frac{i}{2} {\ssl} (\tau_{\mu})_{\alpha}^{\ \beta}{\cal L}_{\beta}, &\\[4pt]
& \{{\cal L}_\alpha,{\cal L}_\beta\}_{\* }=4 
x^\mu\,  (\tau_\mu)_{\alpha\beta} \, ,& \label{osp12b}
\end{eqnarray}
we obtain a representation of the $osp(1|2)$ superalgebra  \eqref{sl2}- \eqref{osp12a}-\eqref{osp12b}. The Casimir operator of the $sl(2,\mathbb{R})\cong so(2,1)$ is equivalent to the hyperboloid constraint \eqref{xx3}
\be \label{Qxx}
x^\mu \* x_\mu = -s(s-1){\ssl} ^2 \,, \qquad s= \frac{1}{4} \, , \frac{3}{4}\,,
\ee
with ``quartion" spins \cite{Volkov:1989qa}. Indeed, the $so(2,1)$ representation spanned by $x^\mu$ consist of the  direct sum of spins  $1/4$ and $3/4$ irreducible representations which are intertwined by the supersymmetry generated by the TPS vectors $\cL_\alpha$. Indeed, the spin Casimir is left invariant
\be \label{xxL}
[x^\mu \* x_\mu , \cL_\alpha]_{\*}= 0\,. 
\ee
The time direction $x^0$ coincides with the expression of the Harmonic oscillator Hamiltonian (cf. \eqref{HHO} and \eqref{xmuqp}). Hence we can use the result for the one-dimensional harmonic oscillator to obtain the correspondent Wigner distributions \eqref{Wn},  ${\ssl}$ now with units of a length scale. Thus the spectrum of the time coordinate reads, 
\be \label{x0ev}
x^0 \* W_n = {\ssl} \Big(\frac{n}{2} +\frac{1}{4} \Big)  \, W_n \, , \qquad n =0,1,2,...  
\ee
It follows that the expectation values of time are
\be \label{x0av}
 \langle x^0 \rangle_n = \int d^2 \cL \; \;  x^0 \* W_n = {\ssl} \Big(\frac{n}{2} +\frac{1}{4} \Big)  \, , \qquad n =0,1,2,...  
\ee
Therefore the  spectrum of the time is positive definite and bounded from below, 
\be \label{x0min}
 \langle x^0 \rangle_0 = \frac{{\ssl}}{ 4} \,, \qquad \langle r^2 \rangle_0 = \frac{{\ssl}^2}{ 4} \,.
\ee

The expectation value of the square of the radial coordinate,
\be \label{r2}
r^2:= x^{1}\* x^{1} +x^{2}\* x^{2} \, ,
\ee
and the time are related through the equation \eqref{xx3}, 
\be \label{r2avgen}
\langle r^2 \rangle =\langle (x^0)^2 + \frac{3}{16} {\ssl}^2 \rangle \, ,  
\ee
hence
\be \label{r2av}
\langle r^2 \rangle_n =\langle (x^0)^2 + \frac{3}{16} {\ssl}^2 \rangle_n = \frac{{\ssl} }{4}(n^2+n+1)  \, , \qquad n =0,1,2,...  
\ee
Thus we obtain the quantization of time and the square of the radius of the fuzzy hyperboloid.
From \eqref{r2av} the area of a disc, defined as 
\be\label{aread}
\langle \mathcal{A}_{disc}\rangle_n= \langle {\pi} r^2 \rangle_n =\frac{\pi {\ssl} }{4}(n^2+n+1)  \, , \qquad n =0,1,2,...  
\ee
is also quantized.

 \subsection{Quantization of the light-cone in $3+1$ dimensions and the Grassmannian $\texttt{Gr}(2,4)$}

 In $3+1$ dimensions, endowing the algebra of multivectors \eqref{X} and \eqref{Z} with the $\*$-product, the deformation of the constraints \eqref{XX}, \eqref{Pluck2} and \eqref{epsiXZ} is given by,
\begin{eqnarray}\label{X*X}
 X^\mu \* X_\mu =\frac{1}{2} {\ssl}^2\,,
\end{eqnarray}
\be \label{*Pluck2}
Z^{[\bmu \bnu} \* Z^{ \blambda \brho]} =0 \,,
\ee
\be \label{epsiX*Z}
 X^{[\bmu } \, \* Z^{\bnu \blambda]} =0 \,.
\ee
Like in the $2+1$ dimensions the light-cone is deformed into a fuzzy hyperboloid, alternatively \eqref{X*X} into a $2+1$ dimensional quantum de Sitter space, while extra directions will appear along the coordinates $Z^{\bmu \bnu}$.   

The quantum equivalent of the constraint \eqref{Pluck2} that defines the Grassmannian subspace $\texttt{Gr}_{\texttt{LC}}(2,4)$ is preserved, hence there exist the notion of \textit{quantum Pluck\"er constraints} and of a fuzzy $\texttt{Gr}_{\texttt{LC}}(2,4)$. 

Let us note here that, more generally, any Lie algebra containing a bi-vector which respect the non-commutative versions of Pl\"ucker constraints can be regarded as containing the quantization of a Grassmannian space $\texttt{Gr}(2,n)$, or some related sub-variety. In particular this is the case of the bi-vectors in references \cite{Snyder:1946qz}, \cite{Heckman:2014xha} where (anti-) de Sitter spaces of $4$ and $5$ dimensions where quantized.

The constraint \eqref{epsiXZ} remains undeformed after quantization, which means that a fuzzy hyperboloid is contained in the plane $Z^{\bmu \bnu}$. 
The identity \eqref{ZZ} changes however,  
\be \label{Z*Z}
 Z^{\bmu \bnu} \* Z_{ \bmu \bnu} = -\frac{3}{2} {\ssl}^2 \,,
\ee
which is related to the change of the norm  of the vectors generating the plane $Z^{\bmu \bnu}$ with respect to the combined Lorentz-scalar product and the $\*$-product. 

With the help of the relations \eqref{LL*LL} and the expressions \eqref{X} and \eqref{Z} we can derive the associative products:
\be \label{asX*X}
X^\bmu \* X^\bnu =X^\bmu  X^\bnu + \frac{i {\ssl}}{2} Z^{\bmu \bnu} +\frac{{\ssl}^2}{8} \eta^{\bmu \bnu} \, ,
\ee
\be \label{asZ*Z}
Z^{\bmu \bnu} \* Z^{ \blambda \brho}  = Z^{\bmu \bnu}  Z^{ \blambda \brho} + i \frac{{\ssl}}{2} (  \eta^{\bnu\blambda}  Z^{\bmu \brho}- \eta^{\bmu\blambda}  Z^{\bnu \brho}+ \eta^{\bnu \brho}  Z^{\blambda \bmu} - \eta^{ \bmu \brho }  Z^{\blambda \bnu} ) + \frac{{\ssl}^2}{8} (\eta^{\bmu \brho}  \eta^{\bnu\blambda} - \eta ^{\bmu\blambda} \eta^{\bnu \brho}    )\,,
\ee
\be \label{asX*Z}
X^\bmu \* Z^{\bnu  \blambda }  = X^{\bmu}  Z^{  \bnu \blambda } -  i \frac{{\ssl}}{2} (\eta^{\blambda \bmu}  X^{ \bnu }- \eta^{\bnu \bmu}  X^{\blambda })\,, \quad Z^{\bnu  \blambda } \* X^\bmu   = X^{\bmu}  Z^{  \bnu \blambda } +  i \frac{{\ssl}}{2} (\eta^{\blambda \bmu}  X^{ \bnu }- \eta^{\bnu \bmu}  X^{\blambda })\,\,.
\ee
Hence,  using the Moyal bracket one finds the following commutation relations,
 \begin{eqnarray}
 [ X^\bmu ,\, X^\bnu ]_\* &=& i {\ssl} Z^{\bmu\bnu}\, ,\label{XX4}\\[4pt]
[ Z^{\bmu\bnu},\, X^{\blambda} ] _\* &=&   i {\ssl}(\eta^{\bnu\blambda}  X^{\bmu}- \eta^{\bmu\blambda}  X^{\bnu})\, ,\label{ZX}\\[4pt]
[ Z^{\bmu\bnu}, \, Z^{\blambda \brho} ]_\*&=& i {\ssl}(  \eta^{\bnu\blambda}  Z^{\bmu \brho}- \eta^{\bmu\blambda}  Z^{\bnu \brho}+ \eta^{\bnu \brho}  Z^{\blambda\bmu} - \eta^{\bmu \brho}  Z^{\blambda\bnu} )\, ,  
 \end{eqnarray}
 thus spanning a representation of  $so(3,2)$ algebra. The operator analogous of this representation is known as harmonic oscillator, metaplectic, or singleton representation  \cite{Flato:1978qz} (introduced by Majorana \cite{Majorana:1932rj,Fradkin:1965zz} but often attributed to  Dirac  \cite{Dirac:1963ta}). Thus to this remarkable $so(3,2)$ representation is associated with the quantiztion of the light-cone and Grassmannian subspace $\texttt{Gr}_{\texttt{LC}}(2,4)$. 

The (anti)commutators,
 \begin{eqnarray}
 [ X^\bmu ,\, \bL_\ba ]_\* &=& -\frac{{\ssl}}{2}(\gamma^\bmu)_\ba{}^\bb \bL_\bb \, , \\[4pt]
[ Z^{\bmu\bnu},\, \bL_\ba ] _\* &=&  - \frac{{\ssl}}{2} \, (\gamma^{\bmu \bnu})_\ba{}^\bb \bL_\bb \, ,\\[4pt]\, 
\{ \bL _\ba ,\, \bL_\bb \}_{\* }&=&2 
X^\bmu\,  (\gamma_\bmu)_{\ba \bb} -  
Z^{\bmu \bnu} \,  (\gamma_{\bmu \bnu})_{\ba \bb}\,,   \label{anticom4}
 \end{eqnarray}
extend the $so(3,2)$ algebra to the $osp(1|4)$ superalgebra. The operator equivalent of this representation of supersymmetry was studied in  \cite{Heidenreich:1982rz,Nicolai:1984gb}. 

The quadratic products \eqref{X*X}, \eqref{*Pluck2}, \eqref{epsiX*Z} and \eqref{Z*Z} are  $osp(1|4)$  invariant, since the infinitesimal transformations generated by $\bL_\ba $ leave invariant the whole set of constraints satisfied by the light-cone and Pl\"ucker  coordinates,
\begin{eqnarray}\label{commallL}
\begin{array}{c}
 [ X^\mu \* X_\mu ,\, \bL_\ba ]_\* = 0 \, , \\[6pt]
[ Z^{[\bmu \bnu} \* Z^{ \blambda \brho]},\, \bL_\ba ] _\* =  0 \, ,\\[6pt]\, 
[ Z^{\bmu \bnu} \* Z_{ \bmu \bnu} , \, \bL_\bb ]_{\* }=0\,\\[6pt]  
[ X^{[\bmu } \* Z^{\bnu \blambda]}  ,\, \bL_\ba ]_\* = 0 \, .
\end{array} 
 \end{eqnarray}

The Wigner function associated to the time $X^0$ \eqref{XXXX} is found by considering its equivalence with the two-dimensional harmonic oscillator Hamiltonian  \eqref{Wn1d}. This is,
\be \label{X0ev}
X^0 \* W_{nm} = \frac{{\ssl}}{2} \, (n+m +1)  \, W_{nm} \, , \qquad n,m =0,1,2,...  
\ee
where
\be
W_{nm}= W_n (q_1,p_1)\* W_m (q_2,p_2) = W_n (q_1,p_1)\, W_m (q_2,p_2)\,, \qquad n,m =0,1,2,...  
\ee
Thus the time expectation values are given by,
\be \label{X0av}
\langle X^0 \rangle_{nm} = \int d^4 \cL \; \;  X^0 \* W_{nm} = \frac{{\ssl}}{2} \, (n+m +1)  \, ,
\ee
The square of the radius of the sphere spanned by the front of light-rays,
\be \label{R2}
R^2:= X^{1}\* X^{1} +X^{2}\* X^{2}+X^{3}\* X^{3} \, ,
\ee
from the constraint \eqref{X*X} equivalent also to
\be \label{R2avgen}
 R^2 = X^0 \* X^0 + \frac{1}{2} {\ssl}^2 \, ,
\ee
is quantized, as it follows from the spectrum of the time,
\be \label{R2av}
\langle R^{2} \rangle_{nm} =  \frac{{\ssl}^2 }{4}\Big((n+m)^2+2(n+m)+3\Big)  \, , \qquad n,m =0,1,2,...   
\ee
Thus their area is given by
\be\label{AS2}
\langle \mathcal{A}_{S^2}\rangle_{nm}= \pi {\ssl}^2 \Big((n+m)^2+2(n+m)+3\Big)  \, , \qquad n,m =0,1,2,...  
\ee
For every fixed time, $X^i$ span fuzzy spheres as positions on their surface cannot be determined exactly, from quantum uncertainty. 

The time and the radius of the fuzzy hyperboloid are bounded from below, 
 \be \label{X0min}
 \langle X^0 \rangle_{00} = \frac{{\ssl}}{ 2} \,, \qquad \langle R^2 \rangle_{00} = \frac{3}{ 4} {\ssl}^2 \,.
\ee

Since $[X^0, Z^3]_\*=0$ \eqref{ZX} the Wigner function for the time coordinate is also a $\*$-eigenfunction of $Z^3$ \eqref{Zi},
\be \label{Z3ef}
Z^{3} \* W_{nm} = \frac{{\ssl}}{2} \, (m-n) W_{nm}\,,
\ee
and hence
 \be \label{Z3av}
\langle Z^{3} \rangle_{nm} =\frac{{\ssl}}{2} \, (m-n)  \, .
\ee

\subsection{Quantization of area and volume}\label{sec:Qareavol}

 Using  the light-cone and Pl\"ucker coordinantes we can construct three independent vectors,
\be \label{rp3basis}
\overrightarrow{X}=(X^1,X^2, X^3)  \,, \qquad \overrightarrow{Y}=(Z^{01},Z^{02}, Z^{03})\,,\qquad \overrightarrow{Z}=(Z^{1},Z^{2}, Z^{3})\,,  
\ee 
which are  orthonormal,
\be\label{aux1}
\overrightarrow{X}\cdot \overrightarrow{X} = \overrightarrow{Y} \cdot \overrightarrow{Y}= \overrightarrow{Z}\cdot \overrightarrow{Z}= (X^0)^2\,, \qquad   \overrightarrow{X} \cdot \overrightarrow{Y} = \overrightarrow{X}\cdot \overrightarrow{Z}= \overrightarrow{Y}\cdot \overrightarrow{Z}=0\,.
\ee 
Here ``$\cdot$" is the scalar product in $\mathbb{R}^3$. By introducing the cross product in $\mathbb{R}^3$  we can verify from the definitions \eqref{X} and \eqref{Z} that,
\be \label{aux2}
 \overrightarrow{X}\times  \overrightarrow{Y} = X^0  \overrightarrow{Z}\,, \quad  \overrightarrow{Y}\times  \overrightarrow{Z} = X^0  \overrightarrow{X} \,, \quad \overrightarrow{Z}\times  \overrightarrow{X} = X^0  \overrightarrow{Y}\,.
\ee 

The products \eqref{aux1} and \eqref{aux2} can be deformed using the  $\*$-product among vectors in addition to the scalar and cross product in $\mathbb{R}^3$, denoted respectively $\cdot_\*$ and $\times_\*$. Then, making use of  the expressions \eqref{asX*X}, \eqref{asZ*Z} and \eqref{asX*Z} we obtain,
\begin{eqnarray}\label{aux1*}
 \begin{array}{c}
\overrightarrow{X}\cdot_\* \overrightarrow{X} = \overrightarrow{Y}\cdot_\*  \overrightarrow{Y}= X^0 \* X^0  + \frac{{\ssl}^2}{2}\,, \qquad  \overrightarrow{Z} \cdot_\* \overrightarrow{Z}= X^0 \* X^0  - \frac{{\ssl}^2}{4} \,,\\[8pt]
 \overrightarrow{X} \cdot_\*  \overrightarrow{Y} = -  \overrightarrow{Y} \cdot_\*  \overrightarrow{X}= -\frac{i3{\ssl}}{2} X^0\,,  \qquad \overrightarrow{X}\cdot_\*  \overrightarrow{Z}= \overrightarrow{Y}\cdot_\*  \overrightarrow{Z}= \overrightarrow{Z} \cdot_\*  \overrightarrow{X}=\overrightarrow{Z}\cdot_\*  \overrightarrow{Y}=0\,,
\end{array}
\end{eqnarray} 
and,
\begin{eqnarray} \label{aux2*}
\begin{array}{c}
 \overrightarrow{X}\times_\*   \overrightarrow{Y} =  X^0 \*  \overrightarrow{Z}\,, \qquad   \overrightarrow{Y}\times_\*   \overrightarrow{Z} = X^0 \*  \overrightarrow{X} - \frac{i3{\ssl}}{2}   \overrightarrow{Y} \, , \qquad   \overrightarrow{Z}\times_\*   \overrightarrow{X} = X^0 \*  \overrightarrow{Y} + \frac{i3{\ssl}}{2}   \overrightarrow{X} 
 \,.
 \end{array}
\end{eqnarray}
From \eqref{aux1*} the orthonormality relations between the vectors \eqref{rp3basis} are deformed. Now the length of the vectors  are
\be\label{Zsphere}
\langle \overrightarrow{Z}\cdot_\*   \overrightarrow{Z}  \rangle_{nm} =  {\ssl}^2 \left(\Big( \frac{n+m}{2}\Big)^2 +  \frac{n+m}{2} \right) \, ,    
\ee
\be\label{XYsphere}
\langle\overrightarrow{X} \cdot_\* \overrightarrow{X} \rangle_{nm} =\langle\overrightarrow{Y} \cdot_\* \overrightarrow{Y} \rangle_{nm} =  \frac{{\ssl}^2 }{4}\Big((n+m)^2+2(n+m)+3\Big)  \, ,   
\ee 
 $n,m =0,1,2,...$ (cf. \eqref{R2av}).

From the cross products  \eqref{aux2*} we obtain a spectrum of the area of the squares with edges ${X}^1$ and ${Y}^2$, i.e.  
$$
 \langle (\overrightarrow{X} \times_\* \overrightarrow{Y})_3\rangle = \frac{{\ssl}^2 }{4} (n+m+1)(n-m)\,.
$$
The volume of the box with edges $(\overrightarrow{X},\overrightarrow{Y},\overrightarrow{Z})$ can be associated to the cyclic combination,
$$
V=\frac{1}3 \Big( \overrightarrow{Z} \cdot_\* (\overrightarrow{X} \times_\* \overrightarrow{Y})+\overrightarrow{X} \cdot_\* (\overrightarrow{Y} \times_\* \overrightarrow{Z})+\overrightarrow{Y} \cdot_\* (\overrightarrow{Z} \times_\* \overrightarrow{X}) \Big)\,,
$$
which insures positive (discrete) expectation values, i.e.
$$
\langle V \rangle_{nm} = \frac{{\ssl}^3 }{8} (n+m+1)\left( (n+m+1)^2+1\right)\,.
$$

\subsection{Quantization of the multivector variety $\bX[\mathcal{M}_D]$,  $D=2,3,4\, ~ \texttt{Mod} ~ 8$} \label{sec:Qn of MV}

In order to construct the non-commutative version of the geometry $\bX[\mathcal{M}_D]$ the juxtaposition product of real numbers should be deformed using  the $\*$-product given in section \ref{PSQ}. 

From \eqref{LL*LL} and the definition of the multivectors \eqref{Xmun} the $\*$-product of multivectors and any function yields in general
\bea \label{X*f1}
\hspace{-0.5cm}\bX^{\bmu[n]} \* f(\bfcL) &=& \bX^{\bmu[n]} f(\bfcL) - i \frac{\ssl}{4} {\ssl} (\Gamma^{\bmu[n]})^{\ba \bb} \bfcL_{(\bfa}  \frac{\partial  f(\bfcL)}{\partial \bfcL^{\bfb)}}  + \frac{{\ssl}^2}{16} (\Gamma^{\bmu[n]})^{\ba \bb} \frac{\partial ^2 f(\bfcL)}{\partial \bfcL^\bfa \partial \bfcL^\bfb}\, ,\\[4pt]
\hspace{-0.5cm} f(\bfcL) \* \bX^{\bmu[n]} &=& \bX^{\bmu[n]} f(\bfcL) + i \frac{\ssl}{4} {\ssl} (\Gamma^{\bmu[n]})^{\ba \bb} \bfcL_{(\bfa}  \frac{\partial f(\bfcL)}{\partial \bfcL^{\bfb)}} + \frac{{\ssl}^2}{16} (\Gamma^{\bmu[n]})^{\ba \bb} \frac{\partial ^2 f(\bfcL)}{\partial \bfcL^\bfa \partial \bfcL^\bfb} \,. \label{X*f2}
\eea
The second line can be obtained from complex conjugation of the first line. Indeed, the complex conjugation, i.e.  changing $i$ by $-i$, in the $\*$-product of two functions, yield and and anti-authomorphism, i.e. $(f\*g)^*=g^* \* f^*$. Thus, complex conjugation is the equivalent of  $\dagger$ operator in quantization in Hilbert space.

We are now interested in the deformation of the constraint \eqref{LVDT}. First of all, using \eqref{X*f1} we obtain
\be \label{hyper2}
\sum _{n=1,2\,\, \texttt{Mod} 4}' \frac{(-1)^{[n/2]}}{n!} \bX^{\bmu[n]} \* \bX_{\bmu[n]} = 2^{2[D/2]-6} (2^{[D/2]}+1)\slashed{\ell}^2_\texttt{H}\,.
\ee
This result reproduces the $2+1D$ case \eqref{xx3} and in $3+1D$ this equivalent to 
$
X^\bmu \* X_\bmu -\frac{1}2 Z^ {\bmu\bnu}\* Z_{\bmu\bnu} = \frac{5}{4}\slashed{\ell}^2_\texttt{H},$ which can be corroborated using \eqref{X*X} and \eqref{Z*Z}.

For a more general result let us define
\be\label{defVDT}
{\bf V_{\* D}}^T:= \left(\begin{array}{c}
\bX^{\bmu} \* \bZ^T_{\bmu} \\[4pt]
\bX^{\bmu[2]}\* \bZ^T_{\bmu[2]}\\
\vdots
\end{array}\right)\,,
\ee 
where $T$ is a multivector parameter entering in the definition \eqref{Znr}. Expanding the components of  \eqref{defVDT} in terms of classical functions we obtain
\bea \label{X*Znr}
\bX^{\bmu[n]} \* {\bZ}^T_{\bmu[n]}= \bX^{\bmu[n]} {\bZ}^T_{\bmu[n]} + \frac{i\ssl}{4} \left( 
 \bar{\bL}\Gamma^{\bmu[n]} \Gamma_{\bmu[n]} T \bL  -  \bar{\bL}\Gamma^{\bmu[n]} T \Gamma_{\bmu[n]} \bL \right)+ \frac{\ssl ^2}{8}\texttt{Tr}~( \Gamma^{\bmu[n]} \Gamma_{\bmu[n]} T) . 
\eea
The latter expression allow us to  obtained the deformed version of  \eqref{XZ},
\bea \label{sumX*Znr}
\bX^{\bmu[n]} \* {\bZ}^T_{\bmu[n]}= &&\sum _{n=1,2\,\, \texttt{Mod} 4}' ({\bf M_D})_n{}^m \left( \bX^{\bmu[m]} \* {\bZ}^T_{\bmu[m]}+  \frac{i\ssl}{4} \bar{\bL}\Gamma^{\bmu[m]} T \Gamma_{\bmu[m]}{\bL} \right)\\[4pt] 
&& + \frac{i\ssl}{4} \left(\frac{3}{2} \texttt{Tr}~( \Gamma^{\bmu[n]} \Gamma_{\bmu[n]} ) \bar{\bL} T \bL -  \bar{\bL}\Gamma^{\bmu[n]} T \Gamma_{\bmu[n]} \bL \right) \\[4pt]
&& + \frac{3}{16}\ssl ^2 \texttt{Tr}~( \Gamma^{\bmu[n]} \Gamma_{\bmu[n]} T) \,,
\eea
and hence of \eqref{LVDT} which takes de form
\be \label{LVDU}
{\bf \Lambda_D} ~ {\bf V_{\* D}}^T- {\bf U_{D}}^T=0\,.
\ee
Thus, quantization sources the original homogeneous equation \eqref{LVDT} with the vector
${\bf U_{D}}^T$, whose entries are
\bea \label{sumX*Znr}
({\bf U_{D}}^T)_n= &&  - \frac{i\ssl}{4}  \sum _{n=1,2\,\, \texttt{Mod} 4}' ({\bf M_D})_n{}^m \, \bar{\bL}\Gamma^{\bmu[m]} T \Gamma_{\bmu[m]} {\bL}\\[4pt] 
&& - \frac{i\ssl}{4} \left(\frac{3}{2} \texttt{Tr}~( \Gamma^{\bmu[n]} \Gamma_{\bmu[n]} ) \bar{\bL} T \bL -  \bar{\bL}\Gamma^{\bmu[n]} T \Gamma_{\bmu[n]} \bL \right) \\[4pt]
&&  - \frac{3}{16}\ssl ^2  \texttt{Tr}~( \Gamma^{\bmu[n]} \Gamma_{\bmu[n]} T) \,.
\eea

The respective traces of products of gamma matrices can be converted in numerical factors using using the identities \eqref{tr} and \eqref{gngmgn}.
Noting that $\bar{\bL} T \bL =\bar{\bL}\Gamma^{\bmu[n]} T \Gamma_{\bmu[n]} \bL =0$ for $T \, \notin \, sp(2^{[D/2]})$ and $\texttt{Tr}~T =0$, and for $T$ non-propotional to the identity we obtain,
we obtain
\be \label{X*ZT1}
\bX^{\bmu[n]} \* {\bZ}^T_{\bmu[n]} =  \bX^{\bmu[n]} {\bZ}^T_{\bmu[n]} \,, \qquad T \, \notin \, sp(2^{[D/2]}) \oplus \mathds{1}\,. 
\ee
So, in these cases, the quadratic relations satisfied by the multivector coordinates remain undeformed in spite of quantization ${\bf \Lambda_D} ~ {\bf V_{\* D}}^T=0$. If $T \propto \mathds{1}$ the star product yields a scalar shift from the classical relations, this is,
\be \label{X*ZT2}
\bX^{\bmu[n]} \* {\bZ}^T_{\bmu[n]}  =  \bX^{\bmu[n]} \* {\bX}_{\bmu[n]} = \bX^{\bmu[n]} {\bX}_{\bmu[n]}+ {\ssl^2} (-1)^{[n/2]}\frac{ D!}{(D-n)! } 2^{[D/2]-3} \,, \qquad T = \mathds{1}/4\,.
\ee

\subsection{Multivector metric in space $\bV[\mathcal{M}_D]$}

Let us define the multivector arrangement
\be \label{hvector}
\bX := \left( \begin{array}{c}
\bX^\bmu \\
\bX^{\bmu[2]} \\
\vdots \\
\bX^{\bmu[n]} \\
\vdots
\end{array} \right)
\ee
where $n=1,2 \texttt{ mod } 4$ and $n \leq D$ for $D$ even, and  $n\leq [D/2]$ for $D$ odd. 

For any two multivectors of the type \eqref{hvector} we define the bilinear forms, with or without $\*$-product activated,
\bea 
{\bf \eta( X,Y)} &=& \sum _{n=1,2\,\, \texttt{Mod} 4}' {\bf \eta_{{\bmu[n]}\, {\bnu[n]} } X^{\bmu[n]} Y^{\bmu[n]}}\,,\label{nXY} \\[5pt]
{\bf \eta_\*( X,Y)} &=& \sum _{n=1,2\,\, \texttt{Mod} 4}' {\bf \eta_{{\bmu[n]}\, {\bnu[n]} } X^{\bmu[n]}\* Y^{\bmu[n]}}   \label{nX*Y}\\[5pt]
\eta_{{\bmu[n]}\, {\bnu[n]} }&=& \frac{(-1)^{[n/2]}}{(n!)^2} \eta_{\bmu_1 \brho_1} \eta_{\bmu_2 \brho_2} \cdots \eta_{\bmu_n \brho_n} \delta_{\bnu_1 \bnu_2 \cdots \bnu_n}^{\brho_1 \brho_2 \cdots \brho_n}. \label{MVmetric}
\eea
The identities \eqref{hyper1} and \eqref{hyper2} can be re-expressed simply as 
\be \label{hyper1'}
{\bf \eta( X, X)}  =  0\,, \qquad {\bf \eta_\*( X , X)} =  2^{2[D/2]-6} (2^{[D/2]}+1)\slashed{\ell}^2_\texttt{H}\,.
\ee
The bilinear form with components $\eta_{{\bmu[n]}\, {\bnu[n]} }$ will be referred as to multivector metric. It is defined in the space of unbounded multivectors $\bV[\mathcal{M}_D]$ \eqref{VD}.

Note that $\eta( X, Y)=\lim_{\ssl\rightarrow 0}\eta_\*( X, Y)$. It is evident that the products \eqref{nXY} and \eqref{nX*Y} are Lorentz invariant. It is less straightforward that with the respective laws of transformation this product will be also $sp(2^{[D/2]})$ invariant. We shall see this in the next section.

The restrictions \eqref{hyper1'} can be regarded as embeddings of a higher dimensional cone, or or its fuzzy deformation, in the linear space $\bV[\mathcal{M}_D]$ of dimension $2^{[D/2]-1} (2^{[D/2]}+1)$. The additional restrictions \eqref{LVDT} define the embedding of the variety $\bX[\mathcal{M}_D]$ in the latter space.

 \subsection{Higher-spin symmetries of $\bX[\mathcal{M}_D]$}\label{HSS}

Quantum oscillators are essential in the construction of higher spin gravity \cite{Vasiliev:1990en,Vasiliev:1992av,Vasiliev:1999ba,Vasiliev:2004cp,Vasiliev:2014vwa}. The theory of Vasiliev extends gauge gravities with generalized spin connections and frame fields which carry not only spin two but all possibles spin. Besides standard statistics \cite{Fradkin:1986ka,Vasiliev:1986qx,Fradkin:1987ah,Konstein:1989ij,Sezgin:2012ag}, in $2+1$ dimensions the Chern-Simons higher spin gravity \cite{Blencowe:1988gj,Vasiliev:1989re} can be extended with fractional spin gauge fields \cite{Boulanger:2013zla,Boulanger:2013naa,Boulanger:2015uha}. The gauge algebras of these theories are referred  as to (fractional) higher spin algebras. The latter are constructed out of Weyl algebras ($\mathfrak{w}$) \eqref{lan} and the star product \eqref{f*g2n}. Vasiliev's higher algebras for $AdS_{D>4}$ spaces \cite{Vasiliev:2004qz}  consists of the quotient $\mathfrak{w}/sl(2)$. Indeed in Vasiliev's theory the amount of oscillator degrees of freedom is proportional to the dimension of  the $AdS$ space. We have observed here that the Weyl algebra of dimension $2^{[D/2]}$ is used to construct multivectors in $D$ dimensions. Thus the amount of oscillator degrees of freedom grows exponentially with the dimension $D$ of the space-time to which it serves as the underlying space for multivectors. A natural question is therefore, whether the multivector space admits higher spin symmetries generated by the whole Weyl algebra, i.e. as symmetries of multivector invariant forms in $D$ space-time dimensions. As we shall see, the answer is yes.  We will still refer as to higher spin symmetries to those generated by the whole Weyl algebra as these generators transform in the adjoint representation of spin-$n/2$ of the Lorentz algebra in $D$ dimensions. 
 
The algebraic constraints satisfied by the coordinates $\bX^{\bmu[n]}$ of the algebraic variety $\bX[\mathcal{M}_D]$, i.e.  \eqref{xx3}, \eqref{X*X}, \eqref{*Pluck2}, \eqref{epsiX*Z}, \eqref{Z*Z}, and \eqref{LVDU}, are consequence of Fierz identities in the algebra of Gamma matrices \eqref{Gamman}, and the triviality of the contraction  $\bfcL^\bfa \, \bfcL_\bfa=0$. These constraints vanish identically, not as imposed. 
Expressing the latter constraints in the homogeneous form
\be\label{C*}
\mathcal{C}_\*(\bfcL)=0\,, 
\ee
and taking the commutator,
\be \label{CL}
[\bfcL_\bfa ,\,  \mathcal{C}_\*(\bfcL) ]_{\*} =  i {\ssl} \, \frac{\partial}{\partial \bfcL^\bfa} \mathcal{C}_\*(\bfcL) =0\,, 
\ee
where we used \eqref{comdif}, means that \eqref{C*} are invariant under infinitesimal TPS transformations.
It follows  from \eqref{CL} that $\mathcal{C}_\*(\bfcL)$ will commute with any powers of $\bfcL_{\ba}$,
\be \label{CLn}
[\bfcL_{\ba(n)}, \mathcal{C}_\*(\bfcL)]_{\*}= 0\,,
\ee
so these constraints are invariant with respect to infinitesimal transformation generated by any monomial $\bfcL_{\ba(n)}$ \eqref{lan}. The algebraic constraints \eqref{C*} will be also invariant also under finite transformations (cf. \eqref{*T}),
\begin{eqnarray}\label{*exp}
   \begin{array}{ll}
g(\varepsilon):=\exp_{\*} (\varepsilon) = \sum_{k=0}^{\infty}\, \frac{1}{k!} & {\underbrace{\varepsilon\* \cdots \* \varepsilon}} \, , \qquad \varepsilon = \sum_{\bfa, n} \varepsilon^{\bfa (n) } \bfcL_{\bfa(n)}\\  
 & k-\hbox{times} \end{array} \,, 
 \end{eqnarray} 
where $\varepsilon^{\ba_1 \cdots \ba_n}$ are constants.
Thus the transformations of functions in phase space,
\be \label{HStransform}
f_g(\bfcL):=g \* f(\bfcL)\* g^{-1}\,, \qquad g \* g^{-1} =1\,,
\ee
leave invariant the the system of constraints \eqref{C*}. Hence the algebraic varieties denoted $\bX[\mathcal{M}_D]$ are higher-spin symmetric. In particular the $sp(2^{[D/2]})$ transformations are generated by the  $\bX^{\bmu[n]}$'s multivectors,
\bea 
\delta \bX^{\bmu[n]} &=& [\varepsilon,\,\bX^{\bmu[n]}]\,,\qquad \varepsilon = \sum'_{m} \varepsilon_{\bnu[m]} \bX^{\bnu[m]}\,, \nonumber\\
&=& \sum'_{m,s}\varepsilon_{\bnu[m]}  C^{\bnu[m] \bmu[n]}{}_ {\blambda[s]} \bX^{\blambda[s]}\,. \nonumber
\eea
Here $\varepsilon_{\bnu[m]}$ are real parameters, $C^{\bnu[m] \bmu[n]}{}_ {\blambda[s]}$ are the constant of structure of the $sp(2^{[D/2]})$ algebra generated by $\bX^{\bmu[n]}$'s and the $\*$-commutator product. The constant of structure can be obtained from the associative algebra of $\Gamma$-matrices and \eqref{X*f1}-\eqref{X*f2}. More generally, for any multivector $\bY^{\bmu[n]}$ in the space $\bV[\mathcal{M}_D]$, the bilinear forms \eqref{nX*Y} are invariant under the infinitesimal $sp(2^{[D/2]})$ transformations,
\bea
\delta \bY^{\bmu[n]} = \sum'_{m,s}\varepsilon_{\bnu[m]}  C^{\bnu[m] \bmu[n]}{}_ {\blambda[s]} \bY^{\blambda[s]}\,.
\eea

{
\section{Multivector matrix models} \label{MVMM}
}

A natural question is whether the multivector space $\bX[\mathcal{M}_D]$ is a solution of any physical model. Since this multivector space consists of non-commutative coordinates and algebraic constraints, it is suggestive to look for relations to matrix models.  
In  \cite{Ishibashi:1996xs} a supersymmetric matrix-model in $0$-dimensions was proposed, the large-$N$ reduced super-Yang-Mills model, which was conjectured to describe type IIB strings in a non-perturbative regime. This is the IKKT model. The authors claim that in this limit the theory should become equivalent to a theory of quantum mechanical operators, on a Hilbert space, which followed by the classical limit should reproduce the Schild action of type IIB strings \cite{Schild:1976vq}. 
Here we would like to show that, a similar matrix model, but  with a fermion mass term and without chemical potential, contains as solutions the non-commutative geometries found so far. Indeed, this model can be obtained using similar arguments than in the IKKT model.

{ Let us consider the Yang-Mills theory with minimally coupled fermions
 \be\label{YM}
S_{YM} = \texttt{Tr}  \left( -\frac{\tilde{\alpha}}{4} \, F^{IJ} F_{IJ} +\frac{1}{2} (\widebar{\psi}\Gamma^I D_I\psi - i m \widebar{\psi} \psi)\right) \,,\qquad I,J=0,1,..., \tilde{D}-1,
\ee
where $D_I\psi = \partial_I \psi + [A_I,\psi]$, is the covariant derivative and $F_{IJ}=\partial_IA_J-\partial_JA_I+A_IA_J-A_JA_I$ is the non-abelian gauge curvature. The formal contraction of the $\tilde{D}$-dimensional space-time,  and trivializing the space-time derivative, $\partial_I=0$, will produce the projection of \eqref{YM} to,}
 \be\label{YM1}
S_{YM} = \texttt{Tr}  \left( -\frac{\tilde{\alpha}}{4} \,[A^I, A^J] [A_I, A_J] +\frac{1}{2} (\widebar{\psi}\Gamma^I [A_I,\psi] - i m \widebar{\psi} \psi)\right) \,,\qquad I,J=0,1,..., \tilde{D}-1.
\ee
Like in the IKKT model  \cite{Ishibashi:1996xs}, we can argue that in the large-$N$ limit the system \eqref{YM1} can be written as a theory of quantum operators. But now, instead of a Hilbert space formalism, we can chose a $\*$-product realization, such that the following model will be obtained,{
\be\label{YM*}
S[\bY,\Psi] =  \left\langle  \left( \, -\frac{{\alpha}}{4} [\bY^I, \bY^J]_\* \* [\bY_I, \bY_J]_\* + \frac{\beta }{2} \widebar{\Psi}\*\Gamma_I[\bY^I,\Psi]_\* - i \frac{\kappa }{2} \widebar{\Psi}\*\Psi \right)\right\rangle \; ,\qquad I,J=0,1,..., \tilde{D}-1,
\ee
}where we relabeled the fields as $\Psi$ and $\bY^I$, and $\alpha$, $\beta$, $\kappa$ are new constants, necessary to produce the correct dimensions of the action, in correspondence with  the dimensions of the fields, $[\Psi]=\ssl^{1/2}$ and $[\bY^I]=\ssl$.
The supertrace operation will be denoted as $\langle \cdot \rangle$.

Here we shall use the supertrace definition  introduced in reference \cite{Vasiliev:1989re} by M. Vasiliev for Heisenberg algebras, which admit Wigner-deformations \cite{Wigner:50}, and their associated universal enveloping algebras. In our case, where the Heisenberg algebras are undeformed, the supertrace operation is reduced to, {
\be \label{str}
\left \langle  f(\bfcL) \right\rangle := f(0), 
\ee
which means that the function $f(\bfcL)$ must be valued in $\bfcL=0$.  As it was shown in reference \cite{Vasiliev:1989re}, the $\mathbb{Z}_2$ degree of these functions under the supertrace, even or odd respectively for $|f|=0,1$, corresponds to the parity of the functions in phase space, 
$$
f(\bfcL)= (-1)^{|f|} f(-\bfcL)\,,
$$
such that, given two functions, under supertrace  they will satisfy the cyclic property,
\be 	\label{cyc}
\left\langle  f(\bfcL)\* g(\bfcL) \right\rangle = (-1)^{|f||g|} \left\langle  g(\bfcL)\* f(\bfcL) \right\rangle \,.
\ee
}For example, 
\be\label{strL*L}
\left \langle  \bfcL_\bfa \* \bfcL_\bfb \right\rangle =i \frac{\ssl}{2} \bfC_{\bfa \bfb}=- \left \langle  \bfcL_\bfb \* \bfcL_\bfa \right\rangle, 
\ee
as it can be deduced from \eqref{L*L4}. { We observe that with respect to the defined supertrace operation the phase space coordinates, i.e. $\bfcL_\bfa$,  have odd degree. This is in some sense remarkable, as at difference of classical Grassmann variables, which are anti-symmetric with respect to the pointwise product, the phase space coordinates pointwise product $\bfcL_\bfa \bfcL_\bfb$ is symmetric, $\bfcL_\bfa \* \bfcL_\bfb$ has a symmetric and an anti-symmetric part, but the supertrace \eqref{strL*L} is anti-symmetric.}

In order to show that the model \eqref{YM*} describes the variety $\bX[\mathcal{M}_D]$, let us identify the labels of fields $I,J$ in \eqref{YM*} with sets of multivector labels,
\be 
I=0,1,2,\dots, \tilde{D}-1 \rightarrow \{0,1,\dots, D\}_1 \oplus \{[01],[02],\dots \}_2\oplus \{[012],[013],\dots \}_3 \oplus \cdots 
\ee
Here, each set $[\bmu]_n:=\{[\bmu_1 \bmu_2\cdots \bmu_n]\,,\; \bmu=0,...,D-1 \}_n$, consists of all independent labels of multivectors of fixed rank-$n$, i.e. ordered increasingly from left to right. To perform the latter correspondence, we need to split the set of indices $ I \,\in \,\{0,1,2,...,\tilde{D}-1\}$ in sub-intervals 
\be 
I\quad  \in \quad \oplus_{n} I_n, \qquad  n = 1,2\,\texttt{ Mod }_4\,,
\ee
such that the elements of the interval $I_n$ will be in one-to-one correspondence with the $\binom{D}{n}$ elements of the set $[\bmu]_n$.  The rank of the multivectors is $n \leq D$ for $D$ even or  $n\leq [D/2]$ for  $D$ odd. Indeed,  $\tilde{D}=2^{[D/2]-1} (2^{[D/2]}+1)$ is the dimension of the space of mutivectors  $\bX[\mathcal{M}_D]$ \eqref{XMD}.
Hence, to every index $I$ we assign an independent multivector index,
 \be \label{idfn}
 I \, \in \, I_{[n]}\quad \rightarrow \quad \bmu[n]_< \, \in \, [\bmu]_n\,,
 \ee
where 
\be \nonumber
  \bmu[n]_< := [\bmu_1\bmu_2\cdots \bmu_n]\,,\quad \bmu_1< \bmu_2 < \cdots < \bmu_n,\quad
 n=1,2~ \texttt{ Mod } 4\,,\quad \bmu=0,1,\cdots D-1\,,
 \ee
represents the label of a multivector increasingly ordered.
Thus we have the following correspondence between single-index fields and multivector fields,
\be 
\bY^I \quad \rightarrow \qquad  \bY^{\bmu[n]_<}\,.
\ee
The signature of the metric $\eta_{IJ}$,  such that $\bY_I =  \bY^I \eta_{IJ}$, is related to the multivector metric \eqref{MVmetric} as follows,
\be 
\texttt{sign}\, \eta_{IJ}=\texttt{sign}\, \eta_{\bmu[n]_< \bnu[n]_<} =(-1)^{[n/2]}\texttt{sign}\,\eta_{\bmu_1 \bnu_1}\eta_{\bmu_2 \bnu_2}\dots \eta_{\bmu_n \bnu_n}\, \,, 
\ee
where the sub-indices have been ordered.
{ Using this new notation we can  rewrite \eqref{YM*} as,
\bea
S[\bY,\Psi] &=&  \left\langle  \left( \, -\frac{{\alpha}}{4} \sum _{m,n=1,2\,\, \texttt{Mod} 4}' \frac{(-1)^{[m/2]+[n/2]}}{m!n!} [\bY^{\bmu[m]},\bY^{\bnu[n]}]_\* \,\*\, [ \bY_{\bmu[m]}, \bY_{\bnu[n]}]_\*  \right. \right. \nonumber \\[5pt]
&& \left. \left. \;\;\;\;\;\;\;\; + \frac{\beta }{2}\sum _{n=1,2\,\, \texttt{Mod} 4}' \frac{(-1)^{[n/2]}}{n!}  \Psi^\a \*(\Gamma_{\bmu[n]})_{\a}{}^\b[\bY^{\bmu[n]},\Psi_\b]_\* -i \frac{\kappa}{2} {\Psi}^\ba\* \Psi_\ba ,  \right)\right\rangle\,,\label{YM*2}
\eea
which is a multivector matrix model in $ D=2,3,4\, ~ \texttt{Mod} ~ 8$ dimensions.
}
The bosonic and fermionic nature of the fields is given by their degree with respect to the supertrace \eqref{str}, $|\bY|=0$, $|\Psi|=1$. For example,
\be 
\langle \bY^{\bmu[n]} \* \bY^{\bnu[m]} \rangle = \langle \bY^{\bnu[m]} \* \bY^{\bmu[n]}  \rangle, \quad \langle \bY^{\bmu[n]} \* \Psi_\bfa \rangle = \langle \Psi_\bfa \* \bY^{\bmu[n]}  \rangle, \quad  \langle \Psi_\bfa \* \Psi_\bfb \rangle = - \langle \Psi_\bfb \* \Psi_\bfa \rangle. 
\ee 

{  \eqref{YM*2} provides the following equations of motion: 
\be\label{EoM1}
\a \sum _{m=1,2\,\, \texttt{Mod} 4}' \frac{(-1)^{[m/2]}}{m!} [\bY^{\bnu[m]}, [  \bY^{\blambda[n]}, \bY_{\bnu[m]}]_\*]_\* +\beta \widebar{\Psi}\*\Gamma^{\blambda[n]} \Psi =0\,,
\ee
from the variation of $\bY$, and
\be\label{EoM2}
\beta \sum _{n=1,2\,\, \texttt{Mod} 4}' \frac{(-1)^{[n/2]}}{n!} (\Gamma_{\bmu[n]})_{\ba}{}^\bb[\bY^{\bmu[n]},\Psi_\bb]_\* - i \kappa  \Psi_\ba =0 \,,
\ee
from the variation of $\Psi$. }
Note that the contraction $\widebar{\psi}^\bfa \* (\Gamma_I)_{\ba}{}^\bb\psi_\bfb= - (\Gamma_I)^{\ba\bb} \widebar{\psi}_\bfa \*\psi_\bfb$, where $ (\Gamma_I)^{\ba\bb}= (\Gamma_I)^{\bb\ba} $,  does not vanish in general. This is in contrast with the classical pointwise product of Grassmann variables (say $\theta_\bfa  \theta_\bfb=-\theta_\bfb  \theta_\bfa$), which is the reason why there are not Majorana currents ($(\Gamma_I)^{\ba\bb} \theta_\bfa  \theta_\bfb=0$). 

These equations of motion admit solutions of the form,
\be \label{sol1}
{  \bY^{\bmu[n]}=\frac{\lambda}{4} \bar{\bL}\Gamma^{\bmu[n]} = \lambda  \bX^{\bmu[n]} , \qquad  \Psi_\ba = b \bfcL_\ba,}
\ee
where $\lambda$ and $b$ are constants to be fixed and $\bX^{\bmu[n]}$ are the multivector coordinates \eqref{Xmun}.
Using \eqref{sol1} we can evaluate the first addend of \eqref{EoM1}, obtaining,
\be\label{Y3}
\sum _{m=1,2\,\, \texttt{Mod} 4}' \frac{(-1)^{[m/2]}}{m!} [\bY^{\bnu[m]}, [ \bY^{\blambda[n]}, \bY_{\bnu[m]}]_\*]_\* =\ssl^2 \lambda^2 2^{[D/2]-1}(2^{[D/2]-1}+1)\bY^{\blambda[n]}\,.
\ee
To prove this we need to use the relation \eqref{LL*LL}, for the $\*$-product of quadratic monomials in $\bfcL$,  and the identity of $\Gamma$-matrices \eqref{Ide1}. When the solution for the fermion field \eqref{sol1} is used in the second addend of \eqref{EoM1} we obtain that,
\be\label{psiGpsi}
\widebar{\Psi}\Gamma^{\blambda[n]} \Psi =\frac{b^2}{\lambda} \bY^{\blambda[n]}\,.
\ee
Both terms, \eqref{Y3} and \eqref{psiGpsi}, when replaced in the equation of motion \eqref{EoM1} produce the following constraint on the constants $b$ and $\lambda$,
\be\label{add1}
 \alpha \ssl^2 2^{[D/2]-1}(2^{[D/2]-1}+1) \lambda^3 +4\beta b^2=0\,.
\ee

When the solutions  \eqref{sol1} are replaced in the equation of motion \eqref{EoM2}, the term $[\bY^{\bmu[n]},\Psi_\bb]_\*$ produces a term proportional to the derivative of $\bY^{\bmu[n]}$ with respect to $\bL^\bb$, which is proportional to  $\bL$. Since the second term in  \eqref{EoM2} is also proportional to $\bL_\ba$, according to \eqref{sol1}, \eqref{EoM2} yields another constraints on the coefficients $\b$, $\lambda$ and $\kappa$,
\be \label{add2}
\ssl \lambda \beta 2^{[D/2]-2}(2^{[D/2]}+1) +\kappa=0\,.
\ee
To obtain \eqref{add2} from \eqref{EoM2} we made also use of the  identities  \eqref{Ide3} and $\bL^\bfa \* \bL_\bfa = i \ssl 2^{[D/2]-1}$. Thus, from \eqref{add1} and \eqref{add2} we obtain the following values of the constant coefficients in \eqref{sol1},
\be \label{constants}
\lambda=-\frac{2^{2-[D/2]}}{2^{[D/2]}+1} \frac{\kappa}{\beta \ssl} \,,\qquad  b=\left(\frac{\a\kappa^3}{2\beta^4 \ssl} \frac{2^{[D/2]-1}+1}{2^{[D/2]}+1}\right)^{1/2}\,.
\ee
Hence the solutions \eqref{sol1} are completely determined. 

{  As consequence of the cyclic property of the supertrace \eqref{cyc}, the action \eqref{YM*}-\eqref{YM*2} enjoys the symmetry \eqref{HStransform},  generated by the Weyl algebra \eqref{*exp}. This is,  one can always move the (even function) $g^{-1}$ in a expression like $\langle g\* f \*g^{-1}\rangle$ to the left $\langle g^{-1}\* g\* f \rangle$, and obtain $\langle g\* f \*g^{-1}\rangle = \langle f \rangle$.
}
This implies that new solutions of the theory \eqref{YM*} can be generated from \eqref{sol1} by means of the transformations   \eqref{HStransform} and parameters \eqref{*exp},
\be \label{sol1f}
\bY'^{\bmu[n]}=g\*\bY^{\bmu[n]}\*g^{-1}, \quad  \Psi'_\ba = g \*\Psi_\ba \* g^{-1}\,.
\ee

{  \subsection{Multivector matrix model in $D=3+1$} }

Let us review the case $D=3+1$ for illustration. The action \eqref{YM*2} take the form:{  
\bea
S[\bY,\Psi] &=&  \left\langle   \, -\frac{{\alpha}}{4}  \Big( [\bY^{\bmu},\bY^{\bnu}]_\* \,\*\, [ \bY_{\bmu }, \bY_{\bnu }]_\* -[ \bY^{\bmu}, \bY^{\bnu\blambda}]_\* \,\*\, [ \bY_{\bmu}, \bY_{\bnu\blambda}]_\* \right. \nonumber \\[5pt]
&&  \;\;\;+\frac{{\alpha}}{4} [ \bY^{\bmu\brho}, \bY^{\bnu\blambda}]_\* \,\*\, [ \bY_{\bmu\brho}, \bY_{\bnu\blambda}]_\* \Big)+ \frac{\beta }{2}\Big(  \Psi^\a \*(\Gamma_{\bmu})_{\a}{}^\b[\bY^{\bmu},\Psi_\b]_\* \label{S4D}\\[5pt]
&& \left. \;\;\; -\frac{1}{2}\Psi^\a \*(\Gamma_{\bmu\bnu})_{\a}{}^\b[\bY^{\bmu\bnu},\Psi_\b]_\* \Big) -i \frac{\kappa}{2} {\Psi}^\ba\* \Psi_\ba ,  \right\rangle\,.\nonumber
\eea
}The equations of motion \eqref{EoM1} are in this case given by,
\bea 
\a [\bY^\bnu,[\bY^\bmu,\bY_\bnu]_\*]_\*- \frac{\a}{2} 
[\bY^{\bnu\blambda},[\bY^\bmu,\bY_{\bnu\blambda}]_\*]_\* + \beta \widebar{\Psi}\*\Gamma^\bmu \Psi &=& 0\,, \label{eom1}\\[5pt]
\a[\bY^\bnu,[\bY^{\bmu\brho},\bY_\bnu]_\*]_\*- \frac{\a}{2} 
[\bY^{\bnu\blambda},[\bY^{\bmu\brho},\bY_{\bnu\blambda}]_\*]_\* +\b \widebar{\Psi} \* \Gamma^{\bmu\brho} \Psi &=& 0\,, \label{eom2}
\eea
and from \eqref{EoM2},
\bea 
\b \; (\Gamma_\bnu)_\ba{}^\bb[\bY^\bnu,\Psi_\bb]_\*- \frac{\b}{2}  (\Gamma_{\bnu\blambda})_\ba{}^\bb[\bY^{\bnu\blambda},\Psi_\bb]_\* -i\kappa \Psi_\ba=0\,. \label{eom3}
\eea
The solutions given in \eqref{sol1}, with the respective parameters \eqref{constants}, now take the specific form :{  
\be \label{sol14D}
\bY^{\bmu}=-\frac{\kappa}{5\beta \ssl} X^\bmu , \quad \bY^{\bmu\bnu}=-\frac{\kappa}{5\beta \ssl} Z^{\bmu\bnu} , \quad \Psi_\ba =\left( \frac{3\a\kappa^3}{250\b^4\ssl}\right)^{1/2} \bfcL_\ba\,,
\ee
}where $X^\bmu$ and $Z^{\bmu\bnu}$  where given in \eqref{X} and \eqref{Z} respectively.
To prove \eqref{eom1} and  \eqref{eom2}  we need to use the $osp(1|4)$ algebra \eqref{XX4}-\eqref{anticom4}. In both cases the bosonic sector exactly compensate the fermionic sector, adding therefore zero. To prove  \eqref{eom3} we need to use \eqref{comdif}, i.e. the property that says that the adjoin action of the operator $[\bfcL_\ba,\cdot]_\*$ yields a derivative with respect to $\bfcL^\ba$, and the  identity,
\be 
(\gamma^\mu \gamma_\mu)_\a{}^\b -\frac{1}{2} (\gamma^{\mu\nu} \gamma_{\mu\nu})_\a{}^\b = 10 \, \delta_\a{}^\b\,,
\ee
which can be obtained from \eqref{Ide3}, for $D=3+1$, or using the explicit representation of the Gamma matrices given below equation \eqref{X}.

Note that the bosonic sector of the multivector matrix model \eqref{S4D} can be expressed  in the form of \eqref{YM*}, i.e. as model in $\tilde{D}=6+4$ dimensions, identifying the  $D=3+1$ multivector indices with $\tilde{D}=6+4$ vector indices \eqref{idfn}. This is,
\be \label{Imun}
I=\{0,1,2,3,4,5,6,7,8,9\} \quad \rightarrow \quad \{\bmu,\bmu[2]\}=\{0,1,2,3,[01],[02],[03],[12],[13],[23]\}\,,
\ee
where $I \in\; I_{[1]} \oplus I_{[2]}$, is decomposed in  $I_{[1]}=\{0,1,2,3\}$ and $I_{[2]}=\{4,5,6,7,8,9\}$, and $[\bmu]_1=\{0,1,2,3\}$ and $[\bmu]_2=\{[01],[02],[03],[12],[13],[23]\}$. Hence  $\bY^I = \bY^\bmu$ for $I=\bmu=0,1,2,3$, $\bY^4 = \bY^{01}$, $\bY^5 = \bY^{02}$, $\bY^6 = \bY^{03}$, $\bY^7 = \bY^{12}$, $\bY^8 = \bY^{13}$, $\bY^9 = \bY^{23}$. And the ten-dimensional metric reads $\texttt{diag }\eta_{IJ}=(-++++++---)$.
This identification yields a model like \eqref{YM*}, which suggest an equivalence with  a $\tilde{D}=6+4$ dimensional Yang-Mills theory (the interaction part). However,  the fermionic term is subtle. In \eqref{S4D} the $\Psi$-field has four components, in agreement with the dimensions of spinors  in $D=3+1$, but  in  the $\tilde{D}=6+4$ perspective the spinor fields must have $32$-components (up to reality conditions). This indicates that the correspondence of the multivector matrix model \eqref{YM*2} in $D=3+1$  with a $\tilde{D}=6+4$  off-shell model \eqref{YM*}, is up to the fermionic terms.  This suggests that to a $\tilde{D}=6+4$ matrix model of the type \eqref{YM*}, with $32$ spinor fermions (matrix) fields, there should be an extended version of the multivector matrix model \eqref{S4D}, in $3+1$ dimensions, with $8$ spinors matrix fields, i.e. 32 fermionic matrix degrees of freedom. The similar conclusions are valid for the most general cases, i.e. for mutivector matrix models  \eqref{YM*2} in $D$-dimensions and matrix models \eqref{YM*} in $\tilde{D}=2^{[D/2]-1} (2^{[D/2]}+1)$. 
Note that there may be pairs of mutivector matrix models in dimensions $D$ and $D\pm1$ dimensions which have associated the same $\tilde{D}$-dimensional matrix model, whenever $[D/2]=[(D\pm1)/2]$.

{
\section{Speculative aspects: Deformation of causal space-time structure and entropy of space-time}\label{NCS}

In this section we explore possible cosmological effects of the quantum deformation of light-cones. We also conjecture on a possible origin of the entropy of space-time, as encoded in the twistor phase space.
}

\subsection{Quantum causal structure}

{ Since light-cones support massless particles, a natural question is how quantum deformations of light-cones may affect their propagation.  As we do not count yet with a theory of particles on non-commutative spaces, even less a field theory, we can just try to look for physical effects of non-commutativity using heuristic arguments.  Here we shall do this.}

Let us  define the {effective hyperboloid},
\be\label{LH}
\mathfrak{r}^{2}= \mathfrak{t}^2 + \ell ^2 \,, \qquad \ft  \,\geq \, \ft_{min}\,,  \quad \fr   \,\geq \, \fr_{min}\,,
\ee
where the radius-square and time-square match the expectation values of $X^0\*X^0$ and $X^i\*X_i$ of the respective non-commutative hyperboloids \eqref{xx3} and \eqref{X*X}, and $\ft_{min}$ and $\fr_{min}$ are the minimal values allowed by the spectrum of the respective observables, i.e.  \eqref{x0av}-\eqref{r2av} and \eqref{X0av}-\eqref{R2av}.  This is,  for $D=2+1$ and $D=3+1$,
\begin{eqnarray}
\hbox{in 2+1 dimensions}&: &   \ell^2=\frac{3}{16}{\ssl}^2 \,, \quad \ft_{min}=\frac{1}{4}{\ssl} ,  \quad \fr^2_{min}=\frac{1}{4}{\ssl}^2 \,, \: \: \label{rmin3} \\[4pt]
\hbox{in 3+1 dimensions} & : & \ell^2=\frac{1}{2}{\ssl}^2 \, ,\quad \ft_{min}=\frac{1}{2}{\ssl} ,  \quad \fr^2_{min}= \frac{3}{4}{\ssl}^2 \,, \: \: \label{rmin4}
\end{eqnarray}
where $\ell$ is obtained from the respective equations \eqref{xx3}-\eqref{X*X}.

From \eqref{LH} one can estimate the average speed of propagation of light,
\begin{eqnarray}\label{coftheta}
c_\ell: = \frac{d \fr} {d\ft}=  \sqrt{1-\frac{\ell^2}{\fr ^2}} \,=\, \sqrt{\frac{\ft^2}{\ft^2+\ell^2}} \,, \qquad \ft  \,\geq \, \ft_{min}\,,  \quad \fr^2  \,\geq \, \fr^2_{min}\, ,
\end{eqnarray}
and its acceleration,
\begin{eqnarray}\label{acc}
a_\ell: = \frac{d^2 \fr} {d\ft ^2}= \frac{\ell^2}{\fr^3} =  \frac{\ell^2}{(\ft^2+\ell^2)^{3/2}}\,, \qquad \ft  \,\geq \, \ft_{min}\,,  \quad \fr^2  \,\geq \, \fr^2_{min}\,.  
\end{eqnarray}
Note that vanishing speed of light and infinite acceleration cannot be observed, since there is a bound at $\fr = \ell$ ($\ft=0$).
We can evaluate the effective speed of light and its acceleration,,
\begin{eqnarray}
\hbox{in 2+1 dimensions}&: &  \begin{array}{l}
 c^\ell_{min}:=\lim_{\fr \rightarrow \fr_{min}} c^\ell 
 =\frac{1}{2} ,  \\[6pt]
  a_{max} :=\lim_{\fr \rightarrow \fr_{min}} a^\ell
  = \frac{3}{2 {\ssl}} \,, \end{array}
\label{ac3}  \\[6pt]
\hbox{in 3+1 dimensions} & : &   \begin{array}{l}
 c^\ell_{min}:=\lim_{\fr \rightarrow \fr_{min}} c^\ell =
 \frac{1}{\sqrt{3}} ,  \\[6pt]
  a_{max} :=\lim_{\fr \rightarrow \fr_{min}} a^\ell 
  = \frac{4}{\sqrt{27}\, {\ssl}}  \,. \end{array} \label{ac4}
\end{eqnarray}
Since in infinity the hyperboloid is asymptotically equivalent to a light-cone, the acceleration $ \lim_{\fr \rightarrow \infty} a^\ell=0$ vanishes.
{ These computations suggest that at small distances light-like signals may look, for a given observer, as propagating with an effective speeds smaller than the classical value $c=1$, but accelerating $a_\ell > 0$. }

\subsection{Bits, entropy and the area of spheres}\label{entro}

The fact that the space $\bX(\mathcal{M}_D)$ makes use of the phase space suggests to implement some notions of statistical mechanics, e.g. distributions and statistical averages already in the classical level. Let us introduce the distribution $\rho(\bfcL)$, such that
\be 
\int \frac{d^{2d}\bfcL}{(2\pi \ssl)^d} \, \rho(\bfcL) = 1\,.
\ee
Note that even-parity distributions can be expressed directly in terms of multivectors $\bX\, \in \, \bX(\mathcal{M}_D)$, i.e. $\rho(\bfcL)=\rho(\bX)$,  from \eqref{LaLb}. The associated entropy functional is given by
\be \label{Gibbscont}
S= - k_\texttt{B} \int \frac{d^{2d}\bfcL}{(2\pi \ssl)^d} \, \rho(\bfcL) \ln \rho(\bfcL)\,,
\ee
 where $k_\texttt{B}$ is the Boltzmann constant and $(2\pi {\ssl})^d$ is the volume of fundamental cell in TPS. 
The discrete version of the Gibbs entropy is given by
 \be \label{Gibbs}
\texttt{S}= - k_\texttt{B}\, \sum_{\texttt{i}}^{\Omega}\, \,  \texttt{p}_\texttt{i} \ln \texttt{p}_\texttt{i} \,,
\ee
where $\texttt{p}_i$ is the inverse of the number of states in the cell $i$, and $\Omega$ is the  total number of cells in the correspondent volume of TPS.

There is an intrinsic probability distribution associated to the degeneracy of the coordinates  $\bX^{\bmu[n]}$ in terms of TPS variables, which is reflected in the symmetry $\bfcL _\bfa \rightarrow - \bfcL_\bfa$ of \eqref{Xmun}. This is, for given values of the coordinates $\bX^{\bmu[n]}$ there are in correspondence two points in TPS, i.e. $\bfcL_\bfa$ and the antipodal $-\bfcL_\bfa$,  which yield the same values of $\bX^{\bmu[n]}$. Therefore the cells $\bfcL_\bfa$ and $-\bfcL_\bfa$ occur with probability $1/2$ each (assuming equal probability weights), whenever the coordinates $\bX^{\bmu[n]}$'s are measured. Hence we can define a bits-state distribution $\rho(\bX)$ in the multivector space, such that it is even with respect to TPS variables, which gives the probability of finding the bit at coordinates $\bX^{\bmu[n]}$ in the state either ``0" or ``1", i.e.  depending if the respective TPS coordinate is found in the  ``lower" or ``upper" TPS half--hyper-plane.  Thus, from \eqref{Gibbs} with $\texttt{p}_\texttt{i}=1/2$ we obtain
\be \label{Gibbs2d}
\texttt{S}= \frac{\ln 2}{2} \, k_\texttt{B}\, {\Omega}\,. 
\ee 
Now consider a solid $2d$-ball in phase space of radius
\be \label{rball}
r= \sqrt{\vec{q}\,{}^2+\vec{p}\,{}^2},
\ee
which has volume
\be\label{vol2d}
V_{2d}=\frac{\pi^d }{d!} r^{2d}\,.
\ee
Hence the number of fundamental unit cells of TPS in this volume is given by  
\be\label{nstates2d}
\Omega = \frac{V_{2d}}{(2\pi {\ssl})^{d}}  \,.
\ee
Replaced in \eqref{Gibbs2d} this yields
\be \label{Gibbs2dX0}
\texttt{S}=k_\texttt{B}\, \frac{\ln 2}{2} \frac{1}{d!} \left(\frac{r^2}{2{\ssl}}\right)^{d}\,.
\ee 
Now, from the definition of the time coordinate \eqref{Xmun} (see also comment above \eqref{X0spec}), 
\be 
r^2=4\bX^0\,,
\ee
the entropy \eqref{Gibbs2dX0} is also equivalent to
\be \label{Gibbs2dX0D}
\texttt{S}= \frac{k_\texttt{B}\, 2^{d-1} {\ln 2}}{d!} \left(\frac{\bX^0}{{\ssl}}\right)^{d}=\frac{k_\texttt{B}\, 2^{(2^{[D/2]-1}-1)} {\ln 2}}{(2^{[D/2]-1})!}  \left(\frac{\bX^0}{{\ssl}}\right)^{2^{[D/2]-1}}\,,
\ee 
where in the second equality we have used the relation $2d=2^{[D/2]}$ between phase space and space-time dimensions. Note that the entropy \eqref{Gibbs2dX0D} increases monotonically with the time,
$$
\frac{d\texttt{S}}{d\bX^0} \geq 0\,. 
$$

{  The entropy \eqref{Gibbs2dX0D} is proportional to the area of a $d$-sphere of radius $R=\bX^0$,}
\be
\mathcal{A}_d= \frac{(d+1)\pi^{(d+1)/2}R^d}{\Gamma[d/2+3/2]}\,
\ee
in spacetime length units. Hence the entropy in TPS can be expressed also in terms of the area of a $d$-sphere,   
\be \label{Gibbs2dXS}
\texttt{S}= \frac{k_\texttt{B}\, (d+2) {\ln 2}}{8 \pi^{d/2} \Gamma[d/2] \ssl^d } \mathcal{A}_d\,.
\ee 
Recalling the fact that the dimension of TPS and the dimension of the space-time are related by the formula $2d=2^{[D/2]}$, in dimensions $D=3$ and $4$ the formulas \eqref{Gibbs2dX0D} and \eqref{Gibbs2dXS} yield an entropy-area relation. 

In more detail, in $D=3+1$ dimensions the entropy of the four-dimensional phase space reads
\be \label{Boltzmann4D}
\texttt{S}= \frac{ k_\texttt{B}\, \ln 2 }{4\pi {\ssl}^2}\mathcal{A}_{2}\,,\qquad D=3+1\,,
\ee
{ which reminds us the Bekenstein-Hawking entropy of a black-hole, or of  cosmological horizons \cite{Gibbons:1976ue,Massar:1999wg,Massar:1999wg,Jacobson:2003wv}). This suggests that twistor (phase) space may provide suitable variables for understanding the origin of the entropy of  black-holes, since it is tempting to compare the entropy/area in TPS \eqref{Boltzmann4D} and the Bekenstein-Hawking entropy,}
\begin{eqnarray}
\texttt{S}_{\texttt{BH}}= \frac{ \mathcal{A}_{BH}}{ 4 \ell_\texttt{P}^2 }\,, \label{BH4d} 
\end{eqnarray}
 where the Planck length is defined as,
\be\label{lp}
\ell_\texttt{P} := \sqrt{ \hbar G_\texttt{N} }\,.
\ee
Comparing \eqref{Boltzmann4D} and \eqref{BH4d} we obtain the proportionality relation between the Heisenberg length and the Planck length,
\begin{eqnarray}
 \ell_\texttt{P} =\sqrt{\frac{ \pi }{\ln 2}} \,\ssl\,, \label{lplH} 
\end{eqnarray}
{ which can be used to express the deformation quantization parameter $\ssl$ in TPS (see \eqref{L*L4}) in terms of the Planck length instead.

Note that in dimensions $D=2+1\,(3+1)$ a constant bit-state distribution yields for the Gibbs/Shannon entropy a perimeter(area)-like Bekenstein-Hawking law. However in $D>4$ the entropy of constant bit-state distributions is proportional to the area of $S^d$-spheres  of dimension greater than the space-time dimensions $D$ (cf. \eqref{dD}). This observation suggest an holographic scenario in $D\leq 4$  in contrast with  $D>4$.

Finally, let us comment on that the existence of lowest bounds for the measurement of time and space distances at  Planck length order of magnitude which together with space-time non-commutativity, implying uncertainty of positions, may overcome the space-time singularities appearing in classical theories of gravity. As discussed in  \cite{Bahns:2015kda}, non-commutativity can prevent also complete evaporation of black holes. Indeed, we can expect that a black-hole should not continue radiating after its event horizon reaches the minimal bound of distance imposed by quantization. Indeed, below this bound the notion of causality does not make sense. In the final state a black-hole must contain just one bit of information. This also suggests a possible particle-like nature of a black-hole as proposed in \cite{Einstein:1938yz,Burinskii:2005mm}. }

  \section{Conclusions}\label{CONC}

 We proposed a formalism that permits a covariantization of phase spaces of dimensions $2d$ as relativistic objects in $D$-dimensional space-time, such that
 \be \label{dD}
 2d=2^{[D/2]}\,.
  \ee
Doing so we extend the notion of Penrose twistor space to phase spaces of dimensions $2^{[D/2]}$ { ($D=2,3,4\, \texttt{Mod}\, 8$) which serve as moduli space of certain multivector varieties, { denoted $\bX[{\cal M}_D]$,} with $D$-dimensional Lorentz covariance. These appear as quadrics in $\mathbb{R}^{\texttt{dim}(sp(2d))}$.} To phase spaces of dimensions \eqref{dD} we referred as to Twistor Phase Space (TPS), emphasizing the fact that they play a dual { role; as non-relativistic mechanical observables in $d$ spatial dimensions, and as spinor-covariant variables in $D$ dimensions.} We observed also that quantum mechanics in phase space produces non-commutative {(fuzzy) versions of $\bX[{\cal M}_D]$} compatible with Lorentz covariance. This approach permits a natural unification of concepts of non-relativistic mechanics, relativistic (algebraic) geometry and (deformation) quantization. 

{ As for physical models underlying these multivector geometries, we found new types of matrix models, which can be regarded as interaction terms of Yang-Mills action principles with massive minimally coupled fermions (in a large-$N$ limit), and whose equations of motion contain as solutions the non-commutative multivector  space $\bX[{\cal M}_D]$.}

{ We included some heuristic arguments on a possible origin of the entropy of space-time and the expansion of the universe. With respect to the first, we observed that a constant bit-state distribution in four dimensional TPS yields a Gibbs/Shannon entropy equivalent to the area of a $2$-sphere embedded in $D=3+1$ dimensions. With respect to the second, we observed that (deformation) quantization in TPS should modify the spacetime causal structure at small scales, since light-cones deforms into fuzzy hyperboloids, with the consequent acceleration of signals. Hence quantum mechanics may source effective acceleration at small scales. }

As it has been shown that  the computation of scattering amplitudes in gauge theory  \cite{Cachazo:2004by,Witten:2003nn,Cachazo:2004zb,ArkaniHamed:2010kv,ArkaniHamed:2010kv,ArkaniHamed:2012nw,Arkani-Hamed:2013kca,Arkani-Hamed:2013jha} can be greatly simplified by means of the use of twistor space and string theories in twistor space \cite{Witten:2003nn,Berkovits:2004hg,Berkovits:2004jj}, { we expect that TPS may be useful to explore generalization} of these results to, perhaps, amplitudes of extended objetcs. We believe also that TPS { may  play a role in the context of Verlinde's entropic-force picture of gravity \cite{Verlinde:2010hp,Brouwer:2016dvq}, as TPS can be easily assigned of an entropy distribution, therefore inducing entropy of geometries immersed in the multivector spaces $\bX[{\cal M}_D]$. } 

\subsection*{Acknowledgments}

I would like to thank  E. Ayon-Beato, N. Boulanger, J. Oliva, P. Ritter, Per Sundell and J. Zanelli, for enlightening discussions. I also acknowledge the kind invitations for long term stays from F. Toppan   (CBPF Institute - Rio de Janeiro) and from Jorge Zanelli (CECs - Valdivia) in their respective institutes. Gratitude also goes to partial support received from FONDECYT postdoctoral grant n$^{\rm o}$ 3120103. Particular thanks go to A. Gorton for helping in the preparation of this manuscript. I would like to thanks also to two referees of this paper for encouraging, firstly, the construction of a physical model (action principle) and, secondly, to provide a more explicit proof of the solutions, among other valuable suggestions. 

\appendix

\section{Gamma matrices}\label{app:Gamma}

The matrices $\Gamma$ satisfy the associative algebra \cite{vanHolten:1982mx},
 \bea 
\Gamma^{\bmu[n]}\Gamma_{\bnu[r]}= \sum_k  \left\{{}^{\bmu[n]}{}_{\bnu[r]} \vert _{\blambda[t_k }{}^{\blambda k-t_k]} \right\} \Gamma^{\blambda [ t_k]}{}_{\blambda [k-t_k]} \, , \\[4pt]
 \left\{{}^{\bmu[n]}{}_{\bnu[r]} \vert _{\blambda[t}{}^{\blambda k-t]} \right\}:= \frac{n! \ r!}{s! \ t! \ u!} \ \delta^{[\bmu_1 \cdots }_{[[\blambda_1 \cdots} \delta^{\bmu_{t} \cdots }_{\blambda_{t} \cdots}\delta^{\bmu_{t+1} \cdots }_{[\bnu_s \cdots}  \delta^{\bmu_{n}] }_{\bnu_1 \cdots} \delta^{\blambda_{t+1} \cdots }_{\bnu_{s+1} \cdots} \delta^{\blambda_{k} ]] }_{\bnu_r]}  \,, \\[4pt]
s:= \frac{(n+r-k)}{2}\,, \qquad  t:= \frac{(n-r+k)}{2}\,, \qquad u:= \frac{(-n+r+k)}{2}\,,
 \eea 
where we have used double square brackets $[[\cdots ]]$ to indicate that $\lambda$-indices are also anti-symmetrized. The sum runs on indices $k$ that does not yield ill-defined result.

The following relations are useful to derive relations of multivectors and solving the equations of motion:
\be \label{Ide3} 
\delta_{\ba}^{\bb} =\frac{1}{2^{[D/2]-1}(2^{[D/2]}+1)} \sum _{m=1,2\,\, ~ \texttt{Mod} ~ 4}' \frac{(-1)^{[m/2]}}{m!} (\Gamma^{\bnu[m]}\Gamma_{\bnu[m]})_{\ba}{}^{\bb}\ ,
\ee

\be \label{} 
\texttt{Tr } (\Gamma^{\bnu[n]}\Gamma_{\bnu[n]}) = - \sum _{m=1,2\,\, ~ \texttt{Mod} ~ 4}' \frac{ (-1)^{[(m+n)/2]} }{ 2^{[D/2]-1} } \frac{n!}{m!} h(D,n,m) \texttt{Tr } (\Gamma^{\bmu[m]}\Gamma_{\bmu[m]}),
\ee

\section{The ${\bf \Lambda_D}$ matrix in $D=9+1,10+1,11+1$ dimensions} \label{app:Lambda}

\be \label{gammalg}
{\bf \Lambda}_{10}=\left(
\begin{array}{cccccc}
 -\frac{5}{4} & -\frac{3}{32} & 0 & \frac{1}{11520} & \frac{1}{1451520} & \frac{1}{11612160} \\[4pt]
 -\frac{27}{16} & -\frac{19}{32} & \frac{1}{384} & -\frac{1}{3840} & -\frac{1}{215040} & \frac{1}{1290240} \\[4pt]
 0 & \frac{105}{2} & -1 & \frac{1}{16} & 0 & \frac{1}{3840} \\[4pt]
 945 & -\frac{315}{2} & \frac{15}{8} & -\frac{15}{16} & \frac{1}{384} & \frac{1}{768} \\[4pt]
 90720 & -34020 & 0 & \frac{63}{2} & -\frac{5}{4} & \frac{1}{32} \\[4pt]
 113400 & 56700 & 945 & \frac{315}{2} & \frac{5}{16} & -\frac{31}{32} \\
\end{array}
\right)\ ,\qquad \texttt{rank}({\bf \Lambda}_{10})=4\,.
\ee

\be 
{\bf \Lambda}_{11}=\left(
\begin{array}{ccc}
 -\frac{41}{32} & -\frac{7}{64} & -\frac{1}{3840} \\[4pt]
 -\frac{35}{16} & -\frac{13}{32} & \frac{1}{384} \\[4pt]
 -\frac{315}{2} & \frac{315}{4} & -\frac{21}{16} \\
\end{array}
\right)\ ,\qquad \texttt{rank}({\bf \Lambda}_{11})=2\,.
\ee

\be 
{\bf \Lambda}_{12}=\left(
\begin{array}{cccccc}
 -\frac{37}{32} & -\frac{1}{16} & -\frac{1}{3840} & 0 & \frac{1}{3870720} & \frac{1}{29030400} \\[4pt]
 -\frac{11}{8} & -\frac{19}{32} & \frac{1}{960} & -\frac{1}{3840} & -\frac{1}{967680} & \frac{13}{58060800} \\[4pt]
 -\frac{495}{2} & 45 & -\frac{21}{16} & 0 & -\frac{1}{5376} & -\frac{1}{40320} \\[4pt]
 0 & -\frac{945}{2} & 0 & -\frac{21}{16} & 0 & -\frac{1}{3840} \\[4pt]
 623700 & -113400 & -\frac{945}{2} & 0 & -\frac{33}{32} & \frac{1}{16} \\[4pt]
 2494800 & 737100 & -1890 & -\frac{945}{2} & \frac{15}{8} & -\frac{19}{32} \\
\end{array}
\right)\ ,\qquad \texttt{rank}({\bf \Lambda}_{12})=4\,.
\ee

\bibliographystyle{utphys}

%
 
\providecommand{\href}[2]{#2}\begingroup\raggedright\endgroup

 \end{document}